\documentclass[a4paper,fleqn,usenatbib,mn2e,article]{mnras}
\usepackage{newtxtext,newtxmath}
\usepackage[T1]{fontenc}
\usepackage{ae,aecompl}
\usepackage{amsfonts, natbib, txfonts,epsfig,rotating,txfonts,mathtools}

\usepackage{graphicx}	
\usepackage{amsmath}	
\usepackage{amssymb}	
\usepackage{pdflscape,tabularx}

\title{The optical vs. mid-infrared spectral properties of 82 Type 1 AGNs: coevolution of 
AGN and starburst}

\author[M. Laki\'{c}evi\'{c} et al.]{
Ma\v{s}a Laki\'{c}evi\'{c},$^{1}$\thanks{E-mail: mlakicevic@aob.rs}
Jelena Kova\v{c}evi\'{c}-Doj\v{c}inovi\'{c},$^{1,2}$
\& Luka \v{C}. Popovi\'{c}$^{1,2}$
\\
$^{1}$Astronomska Observatorija Beograd; Volgina 7, 11000 Beograd, Serbia\\
$^{2}$Isaac Newton Institute of Chile, Yugoslavia Branch\\
}

\pubyear{2017}

\begin{document}
\label{firstpage}
\pagerange{\pageref{firstpage}--\pageref{lastpage}}
\maketitle

\begin{abstract}

We investigated the connection between the mid-infrared (MIR) and optical spectral characteristics 
in a sample of 82 Type 1 active galactic nuclei (AGNs), observed with Infrared Spectrometer on Spitzer (IRS) 
and Sloan Digital Sky Survey (SDSS, DR12). We found several interesting correlations between optical and MIR 
spectral properties: i) as starburst significators in MIR increase, the EWs of 
optical lines H$\beta$NLR and FeII, increase as well; ii) as MIR spectral index increases, 
EW([OIII]) decreases, while fractional contribution of AGN (RAGN) is not connected with EW([OIII]); 
iii) The log([OIII]5007/$\rm H\beta$NLR) ratio is 
weakly related to the fractional contribution of polycyclic aromatic hydrocarbons (RPAHs). We compare the two 
different MIR and optical diagnostics for starburst contribution to the 
overall radiation (RPAH and BPT diagram, respectively). The significant differences between optical and MIR 
starburst diagnostics were found. The starburst influence to observed correlations between
optical and MIR parameters is discussed. 

\end{abstract}


\section{Introduction} \label{sec:intro}

Understanding the nature of coexistence of active galactic nuclei (AGN)
and surrounding starburst (SB) is one of the main problems of galactic evolution.
Coexistence of AGNs and SBs is found in various samples such as 
hyperluminous infrared (IR) galaxies \citep{Ruiz13}, ultra(luminous) IR galaxies \citep{Kirkpatrick15}, 
Seyfert 1 and Seyfert 2 galaxies \citep{Dixon11}, studied in models \citep{Lipari06} and often discussed
in the frame of AGN spectral properties \citep{Mao09,Sani10,Popovic11,Feltre13,Melnick15,Ishibashi16,Contini16}. There are some indications that AGNs may 
suppress star formation and gas cooling \citep{Croton06,man16}, and
that star formation is higher in AGNs with lower black hole (BH) mass \citep{Sani10}. Nevertheless, some studies show a 
correlation between star formation rate and AGN luminosity, at high luminosity AGNs \citep{Lutz10,Bonfield11}. 

To find SB contribution to the AGN emission there are several methods in the optical and MIR spectrum.
The empirical separation between the low-ionization nuclear emission-line regions (LINERs), HII regions 
and AGNs at the optical wavelengths is the BPT diagram \citep{Baldwin81,Kewley01,Kauffmann03}, usually given 
as the plot of the flux ratio of forbidden and allowed narrow lines log($\rm [NII]6563/H\alpha NLR$) vs. 
log($\rm [OIII]5007/H\beta NLR$). The main diagnostic assumption 
is that HII regions are ionized by young massive stars, while AGNs are ionized by high energetic
photons emitted from the accretion disc. In the case that the H$\alpha$ spectral range is not present 
in the AGN spectra, the R=log([OIII]5007/$\rm H\beta$NLR) ratio may indicate the contribution of SB to the AGN 
emission, as e.g. \citet{Popovic11} suggested that the dominant SBs have R$<$0.5, 
while dominant AGNs have R$>$0.5. 

Mid-infrared (MIR) based probes for the star formation suffer much less from the extinction than ultraviolet, optical and near-IR observations. The first comparison between Type 1 and Type 2 AGNs at 2-11 $\mu$m range is done by \citet{Clavel00}. The star-forming galaxies are expected to have stronger polycyclic aromatic hydrocarbon (PAH) features \citep{ForsterSchreiber04,Peeters04,Wu09}. \citet{ForsterSchreiber04} tested the relation between both 5-8.5$\mu$m PAH and 15$\mu$m continuum emission with Lyman continuum and found that MIR dust emission is a good tracer for the star formation rate. \citet{Shipley16} used the H$\alpha$ emission to calibrate PAH luminosity as a measure of a star formation rate.   

\citet{Brandl06} and \citet{Dixon11} discussed the discrimination of SBs from AGNs using the 
$\rm F_{15 \mu m}/F_{30 \mu m}$ continuum flux ratio. SBs have steeper MIR spectrum since AGNs produce
a warm dust component in MIR \citep{Dixon11}. The $\rm F_{15 \mu m}/F_{30 \mu m}$ ratio measures
the strength of the warm dust component in galaxies and therefore reflects SB/AGN contribution. 
\citet{Veilleux09} have similar conclusion, that SB galaxies have 
log($\rm F_{\rm 30 \mu m}/F_{\rm 15 \mu m}$) of 1.55, while AGNs
have log($\rm F_{\rm 30 \mu m}/F_{\rm 15 \mu m}$) of 0.2. As a consequence, 6.2$\mu$m 
PAH equivalent widths (EWs) correlate with 20-30$\mu$m spectral 
index \citep{Deo07}. A lower 25 to 60 $\mu$m flux ratio in SB galaxies than in Seyfert galaxies is 
explained by cooler dust temperature in SB galaxies. The higher PAH EWs in SB galaxies than in Seyfert 
galaxies is due to the PAH destruction from the high energetic radiation from the AGN accretion disc 
\citep{Wu09}. Another reason for PAH absence in 
luminous AGNs is a strong MIR continuum that can wash-out the PAH features, reducing
their EWs \citep{Laurent00,AlonsoHer14}.

MIR lines that could indicate an AGN presence are [NeV]14.32, [NeV]24.3, [SIV]10.51$\mu$m
\citep{Spinoglio92,Dixon11,Chen14}, [OIV]25.9$\mu$m can originate both from starforming regions or 
AGNs \citep{Lutz98a,Wu09}, while nearby [FeII]25.99$\mu$m is primarily due star formation 
\citep{Hartigan04,Weedman05}. [NeII]12.8$\mu$m is strong in SB galaxies, but weak in AGNs \citep{Dixon11}. 
Therefore, \citet{Genzel98,Lutz98a,Deo07} and \citet{Dixon11} used various methods 
to distinguish AGNs from SBs: [NeIII]15.6$\mu$m/[NeII]12.8$\mu$m, [NeV]14.3$\mu$m/[NeII]12.8$\mu$m, 
[NeV]24.3$\mu$m/[NeII]12.8$\mu$m, [OIV]25.9$\mu$m/[NeII]12.8$\mu$m, [OIV]25.9/[SIII]33.48 ratios, as well 
as a fit to the MIR spectral-slope and strength of the PAH features. 

There is a number of correlations between the UV, optical and IR spectral properties that can be
caused by physical characteristics of AGNs, but also by contribution of SB to the AGN emission. Some of
these correlations can be an indicator of SB and AGN coevolution. Note here shortly the
results of \citet{Boroson92} (BG92 afterwards), who performed principal component analysis (PCA) on various 
AGN optical, radio and X-ray characteristics. They found a set of correlations between different spectral 
parameters, projected to eigenvector 1 (EV1). Some of these are anticorrelations EW([OIII]) vs. 
EW(FeII) and full width at half maximum (FWHM) of $\rm H\beta_{broad}$ vs. EW(FeII) 
at optical wavelengths. These correlations in AGN spectral properties are very intriguing and their 
physical background is not understood. Other authors performed PCA in different AGN samples in order 
to explain these correlations \citep{Grupe04,Mao09}. \citet{Wang06} and \citet{Popovic11} suggested that the 
SB affects EV1, while \citet{Feng15b} found that PAH characteristics are correlated with EV1. \citet{Bian16} 
performed PCA on Spitzer spectra of QSOs and compared these results with the ones from BG92.

Many authors have compared the optical and MIR observations of AGNs 
\citep[see e.g.][]{Sajina08,Sani10,Lacy13,Vika17}. \citet{Odawd09} and \citet{LaMassa12} compared the BPT diagram classification 
and MIR AGN/SB diagnostics and concluded that the congruence is high. On the other hand, \citet{Vika17} 
showed that the optical BPT AGN classification does not always match the one obtained from spectral energy 
distribution (SED) fitting from UV to far-infrared (FIR) wavelengths. \citet{Goulding09} and \citet{Dixon11} 
used Spitzer data and 
found that there may exist AGNs in half of the luminous IR galaxies without any evidence of AGN at 
near-infrared and optical wavelengths, undetected because of the extinction.

At the MIR wavelength range one can see hot, AGN heated dust component, from the pc-size region 
surrounding the central BH, that could be the reservoir that feeds the central BH during the 
accretion phase \citep{Feltre13}. At the optical observations of Type 1 AGNs, the contributions 
of accretion disc and broad line region are seen. However, the optical and MIR emission of an AGN should be 
related, as e.g. if the emission of central continuum source is stronger, one can expect that the inner
part of torus is larger \citep{Barvainis87}. Therefore, the correlations between
MIR and optical emission in these objects were expected and found \citep{Shao13,Singal16}.

In this work we use optical (SDSS) and MIR (Spitzer) data to investigate the correlations
between optical and MIR emission of Type 1 AGNs, and compare the influence of SB/AGN to the optical and MIR 
spectra. The paper is organized as follows: in Section~\ref{sec:sample}, we describe our sample 
of Type 1 AGNs, in Section~\ref{sec:Analysis} we explain our data analysis for optical and MIR, 
in Section~\ref{sec:results} we present the results, in Section~\ref{sec:discuss} we discuss our 
results, and in Section~\ref{sec:conclussion} we outline our conclusions.

\section{The sample} \label{sec:sample}

\subsection{The sample of Type 1 AGNs}
In this research, we used the sample of Type 1 AGNs, found in the cross-match 
between optical Sloan Digital Sky Survey (SDSS) spectra and MIR Spitzer Space Telescope spectral data. 

The SDSS Data Release 12 (DR12) \citep{2015ApJS...219..12A} contains all SDSS observations 
until July 2014. These observations were done with 2.5 m telescope at the Apache Point Observatory with 
two optical spectrograph (SDSS-I and BOSS). All data from prior data releases are included in DR12 
and re-analyzed, so that it contains in total 477,161 QSO and 2,401,952 galactic optical spectra. The SDSS-I 
spectra covering the wavelength range of 3800 \AA \ to 9200 \AA \ with spectral resolution 1850--2200, 
while BOSS spectra are observed in the range 3650-10400 \AA, with spectral resolution of 1560--2650.

IR data used in this work are reduced and calibrated 5-35 $\mu$m IRS\footnote{The Infra-red 
Spectrograph \citep[IRS;][]{Houck04} on-board the Spitzer Space telescope.} spectra, available in the 
$\rm 6^{th}$ version of The Cornell Atlas of Spitzer/IRS Sources 
(CASSIS\footnote{The spectra are taken from CASSIS (also known as the Combined Atlas of Sources with Spitzer IRS 
Spectra) web-page: \url{http://cassis.sirtf.com/atlas/}.}) database, in 
low-resolution of $\rm R\sim 60-127$ \citep{Houck04,Werner04,Lebouteiller10,Lebouteiller11}. 

To find the sample for this investigation, we used Structural Query Language (SQL) to search for 
all galaxies and quasars in SDSS DR12 which satisfy following criteria: 
\begin{enumerate}
\item $S/N>$15 in g--band (4686{\AA}), in order to obtain the spectra of an adequate quality for fitting procedure,
\item $z<0.7$, with $z_{warning}$=0 matching the spectra which cover the optical range near the H$\beta$ line,
\item the objects are classified as 'QSO' or 'galaxy' in SDSS spectral classification.
\end{enumerate}

\begin{figure} 
 \centering
 \includegraphics[width=96mm]{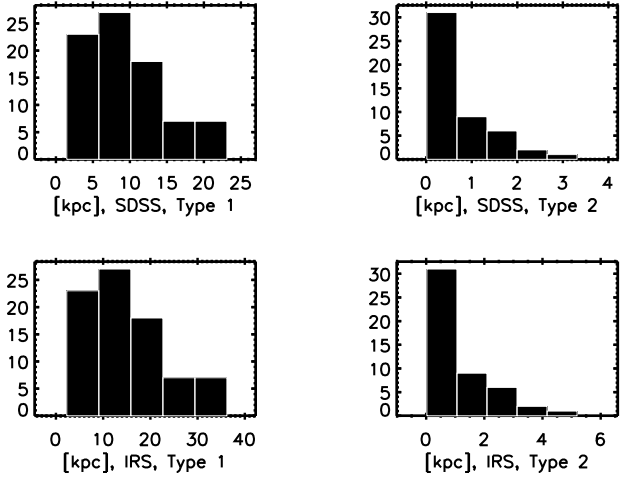}
\caption{The distributions of the SDSS and IRS aperture projections for the Type 1 AGN sample are given on the left, while the distributions of the SDSS and IRS aperture projections for the Type 2 AGN sample are given on the right. \label{fig:apert}}
 \end{figure}

 The resulting search contained 135,633 objects. These objects were cross-matched in TOPCAT 
 with the latest (from November 2015) Spitzer catalog -- Infrared Database of Extragalactic 
 Observables from Spitzer (IDEOS\footnote{\url{http://ideos.astro.cornell.edu/redshifts.html.}}) of 
3361 extragalactic sources \citep{2016MNRAS...445..1796}. This cross-match resulted with 585 objects. 
From that sample we selected Type 1 AGN by visual 
inspection of the optical spectra, removing all objects which do not have broad emission lines 
in the $\lambda\lambda$4000-5500 {\AA} range. That resulted with 98 Type 1 AGNs. Additionally, 16 objects 
were removed from the sample because of high noise and/or
poor fitting of IR spectra (see Section~\ref{MIR_properties}). Finally, the sample contains 82 AGN 
Type 1, with optical and MIR spectra of satisfactory quality (see Table~\ref{tab:Sy1}). The angular
size distances are given in the column (7), while the projected linear diameters of the IRS and SDSS 
apertures are given in the columns (8) and (9). The distributions of the SDSS and IRS aperture projections for Type 1 sample are given in the Fig.~\ref{fig:apert}, on the two panels -- left. The SDSS aperture size is 3$^{\prime\prime}$ \citep{Rosario13}, while we took 4.7$^{\prime\prime}$ for the IRS aperture \citep{Houck04}.

\begin{table*}
\caption{The sample of Type 1 AGNs, observed at the MIR and optical wavelengths. A full version is available at 
the electronic format. \label{tab:Sy1}}
\centering{
\begin{tabular}{lcclccccc}
\hline\hline
NED name& RA[$^{\circ}$] & Dec[$^{\circ}$] & SDSS name & plate-MJD-fiber &Z &Distance [Mpc] &SDSSProjA [kpc]& IRSProjA [kpc]\\
(1) &(2)&(3)&(4)&(5)&(6)&(7)&(8)&(9) \\
\hline  
  MRK\_0042     &    178.42407& 46.21174& J115341.78+461242.25& 1446-53080-0129& 0.0243 &101.0& 1.47&2.30\\
  SBS\_1301+540 &    195.99791& 53.79165& J130359.48+534729.8 & 6760-56425-0390& 0.0301 &124.3& 1.81&2.83\\ 
  MRK\_0290     &    233.96835& 57.90264& J153552.40+575409.50& 0615-52347-0108& 0.0302 &124.7& 1.81&2.84\\
  UM\_614       &    207.47018& 2.07919 & J134952.84+020445.10& 0530-52026-0165& 0.0329 &135.4& 1.97&3.08\\
  MRK\_0836     &    224.75565& 61.23155& J145901.36+611353.59& 0611-52055-0437& 0.0389 &158.9& 2.31&3.62\\   
\hline
\end{tabular}}
\\
\smallskip
{\bf Notes.}
In the column (1) we give NED name, in the columns (2) and (3) -- Right Ascention and the Declination in
the arc degrees, in the columns (4) and (5) -- SDSS name and plate-MJD-fiber numbers, respectively. In the
column (6) is the redshift, column (7) -- Angular size distance, calculated using 
\url{http://www.astro.ucla.edu/\~Ewright/CosmoCalc.html}, column (8) -- Projected SDSS aperture diameter and 
column (9) -- Projected IRS aperture diameter. 
\end{table*}

\subsection{The sample of Type 2 AGNs from the literature} \label{mixed}

Because of the difficulties in decomposition of the narrow H$\alpha$, H$\beta$ and [NII] lines in 
the initial sample of Type 1 AGNs, to check some relations between narrow lines, we also consider 
a sample of Type 2 AGNs, taken from the literature (more in Section~\ref{sec:CompareBPT}). The sample 
is taken from \citet{2015ApJ...803..109H} and \citet{Garcia16}, 
as these authors performed the fitting of CASSIS spectra of $\sim$150 objects -- Seyfert 2 
galaxies, LINERs and HII regions, using {\sl deblendIRS} routine (we use the same routine in
this work). We took the fitting results that they obtained (spectral index, $\alpha$ and fractional 
contributions of AGN and PAH in the spectra -- RAGN and RPAH, respectively) for all objects for which we 
found the log([OIII]/H$\beta$) and log([NII]/H$\alpha$) measurements in the other literature. The sample has 
49 Type 2 AGNs and the data that we use are given in the Table~\ref{tab:lit}. The angular
size distances are given in the column (9), while the projected linear diameters of the IRS and SDSS 
apertures are given in the columns (10) and (11). The distribution of these aperture projections is given on 
the Fig.~\ref{fig:apert}, on the two panels -- right.

\begin{table*}
\caption{The sample of Type 2 AGNs taken from the literature and the data that have been used in this work. A 
full version is available at the electronic format. \label{tab:lit}}
\centering{
\begin{tabular}{ccccccccccc}
\hline\hline
ID & $\alpha$ & RPAH & RAGN & log([OIII]/H$\beta$) & log([NII]/H$\alpha$) & Refopt & RefIR & Distance [Mpc] & SDSSProjA [kpc]& IRSProjA [kpc]\\
(1) &(2)&(3)&(4)&(5)&(6)&(7)&(8)&(9)&(10)&(11)\\
\hline
NGC1052       &-2.46&0.053&0.787 &0.303 &0.079 &Ho97      &Her15 & 21.283 & 0.019 & 0.029\\
NGC1275       &-2.80&0.0  &1.000 &1.173 &0.134 &Ho97      &Her15 & 73.778 & 1.073 & 1.681\\
NGC1386       &-2.18&0.130&0.734 &1.570 &0.200 &Kewle01   &Her15 & 12.376 & 0.180 & 0.282\\
NGC1614       &-2.54&0.702&0.298 &-0.090&-0.220&Kewle01   &Her15 & 66.789 & 0.971 & 1.521\\
NGC1808       &-1.66&0.949&0.047 &-0.796&-0.268&Gould09   &Her15 & 14.076 & 0.205 & 0.321\\
\hline
\end{tabular}}
\\
\smallskip
{\bf Notes.}
In column (1) are object names, in columns (2)-(4) are MIR measurements from the 
literature, in columns (5)-(6) are optical spectroscopic data from the literature, column (7) -- 
reference for the optical data, column (8) -- reference for the MIR data, column (9) -- 
Angular size distance, calculated using \url{http://www.astro.ucla.edu/\~Ewright/CosmoCalc.html}, column 
(10) -- Projected SDSS aperture diameter and column (11) -- Projected IRS aperture diameter. Feng15b=\citet{Feng15a}, Gar16=\citet{Garcia16}, Haas07=\citet{Haas07}, 
Her15=\citet{2015ApJ...803..109H}, Ho97=\citet{Ho97}, Kewle01=\citet{Kewley01}, 
LaMassa=\citet{LaMassa11}, Gould09=\citet{Goulding09}, Shi10=\citet{Shields90}, Veilleux95=\citet{Veilleux95}.
\end{table*}

\section{Analysis} \label{sec:Analysis}

\subsection{AGN optical properties}

The SDSS spectra were corrected for the Galactic extinction by using the standard Galactic-type 
law \citep{ho1983} for optical-IR range and Galactic extinction coefficients given by \citet{Schlegel98}, 
available from the NASA/IPAC Infrared Science Archive 
(IRSA)\footnote{\url{http://irsa.ipac.caltech.edu/applications/DUST/}}.
Afterwards, the spectra were corrected for cosmological redshift and host galaxy contribution 
(Section~\ref{host_sub}), and fitted using model of optical emission 
in $\lambda\lambda$4000-5500 {\AA} and $\lambda\lambda$6200-6950 {\AA} ranges (Section~\ref{model_line}).

\subsubsection{Host galaxy subtraction} \label{host_sub}

To determine the host galaxy contribution in the optical spectra we applied the 
PCA \citep{1992ApJ...398..476, 2006AJ...131..84}. PCA is a statistical method 
which enables a large amount of data to be decomposed and compressed into independent components. In the case of the spectral PCA, these 
independent components are eigenspectra, whose linear combination can reproduce the observed spectrum. 
Spectral principal component analysis is commonly used for classification of galaxies and QSOs 
\citep[etc.]{1992ApJ...398..476, 1995AJ...110..1071, yip04a, yip04b}. \citet{yip04a} used 170,000 galaxy 
SDSS spectra to derive the set of the galaxy eigenspectra, while \citet{yip04b} used 16,707 QSO SDSS spectra 
to construct several sets of eigenspectra which describe the QSO sample in different redshift and luminosity 
bins.

 \citet{2006AJ...131..84} introduced the application of this method for spectral decomposition into pure-host 
 and pure-QSO part of an AGN spectrum. This technique assumes that the composite, observed spectrum can be 
 reproduced well by the linear combination of two independent sets of eigenspectra derived from the 
 pure-galaxy and pure-quasar samples. They applied different sets of galaxy and QSO eignespectra derived 
 in \citet{yip04a,yip04b}, and 
 found that the first few  galaxy and quasar eigenspectra can reasonably recover the properties of the sample.

Following the procedure described in \citet{2006AJ...131..84}, we used the first 10 QSO eigenspectra 
derived from high-luminosity (C1), low-redshift range (ZBIN 1), defined by \citet{yip04b}, and the 
first 5 galaxy eigenspectra derived in \citet{yip04a}. The galaxy eigenspectra are downloaded from 
SDSS Web site\footnote{\url{http://classic.sdss.org/dr2/products/value\_added/}}, while the QSO eigenspectra 
are obtained in the private communication \citep{Yippriv}.

 \begin{figure} 
 \centering
 \includegraphics[width=96mm]{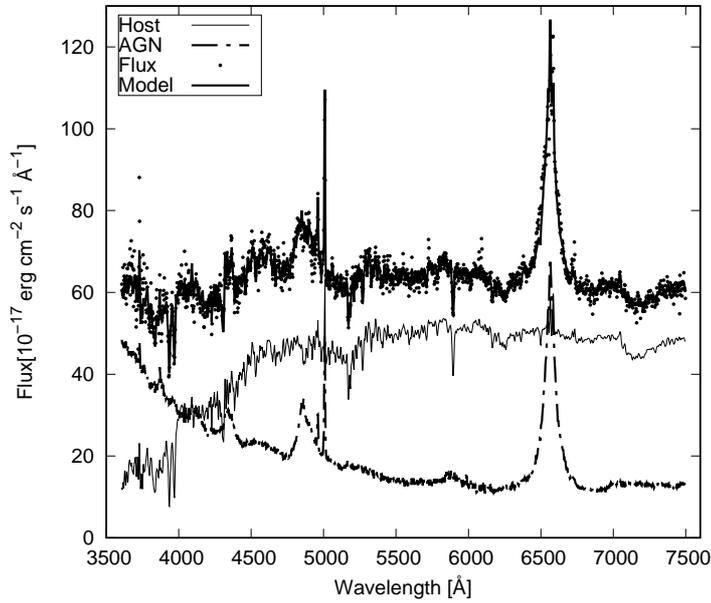}
\caption{An example of the spectral PCA decomposition of object SDSS\,J163631.28$+$420242.5 (0628-52083-0302) 
on the pure host and the pure QSO part.   \label{fig:J1}}
 \end{figure}

 \begin{figure} 
 \centering
 \includegraphics[width=96mm]{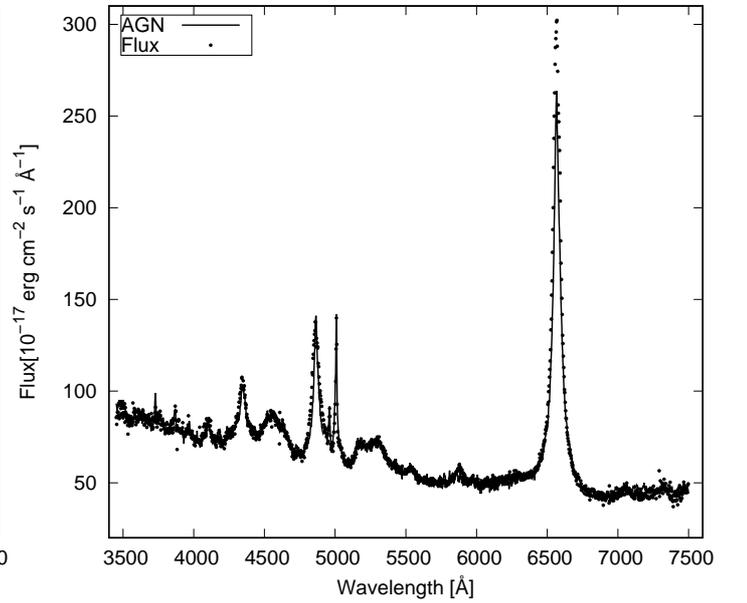}
\caption{An example of the spectral PCA decomposition of object J005812.85+160201.37 (0421-51821-0329) 
which we fitted only with QSO eigenvectors, without galaxy eigenvectors.   \label{fig:nohost}}
 \end{figure}

First, we re-bined the observed spectra and 15 eigenspectra to have the same range and wavelength bins. 
Afterwards, we fitted our spectra with a linear combination of QSO and galaxy eigenspectra. 
The host part of the spectrum is derived as a linear combination of galaxy, while the AGN part
as a linear combination of QSO eigenspectra. We masked all 
narrow emission lines from the host galaxy part, and subtract only the host galaxy continuum and stellar 
absorption lines from the observed spectra. Finally, we performed the fitting on that spectra. Our model is 
described in the Section~\ref{model_line}, while measuring of optical parameters is explained in 
Section~\ref{measuring_opt}. The example of spectral PCA 
decomposition is shown in Fig.~\ref{fig:J1}. The host fraction, F$_{H}$ is measured in the
$\lambda\lambda$4160-4210{\AA} range, and presented in Table~\ref{tab:opt}.

In some cases, the fitting results may give a non-physical solutions, such as negative host or AGN part. 
\citet{2006AJ...131..84} also had non-physical solutions in their Table 2. In our sample we had several
objects where fit gives the negative host contribution. In these cases, we assumed that the host part is 
equal to zero, therefore we excluded galaxy eigenvectors from our 
fitting and fitted only with the QSO eigenvectors, since we find that these fits are usually good in 
the part of the spectra we were interested in (see Fig.~\ref{fig:nohost}).

\subsubsection{Model of the line spectra in the optical range} \label{model_line}

After the host galaxy contribution is subtracted from the observed spectra, the power-law 
SED, typical for the quasars, with the broad and narrow emission lines is obtained. The QSO 
continuum is estimated using the continuum windows given in \citet{ku2002}. The points of the continuum 
level are interpolated and the continuum is subtracted \citep[see][]{Kovacevic10}.
 
The optical emission lines were fitted in two ranges: $\lambda\lambda$4000-5500 {\AA} in order to cover Balmer lines (H$\beta$, H$\gamma$, H$\delta$), optical FeII 
and [OIII] $\lambda\lambda$4959, 5007{\AA} lines, and $\lambda\lambda$6200-6950 {\AA}, 
where H$\alpha$ and [NII] $\lambda\lambda$6548, 6583{\AA} lines are present. For fitting 
the emission lines we used a model of multi-Gaussian functions \citep{po2004}, where each Gaussian is assumed 
to represent an emission from one emission 
region \citep[see][and references therein.]{Kovacevic10,Kovacevic15}

\begin{figure*} 
 \centering
 \includegraphics[width=109mm,angle=270]{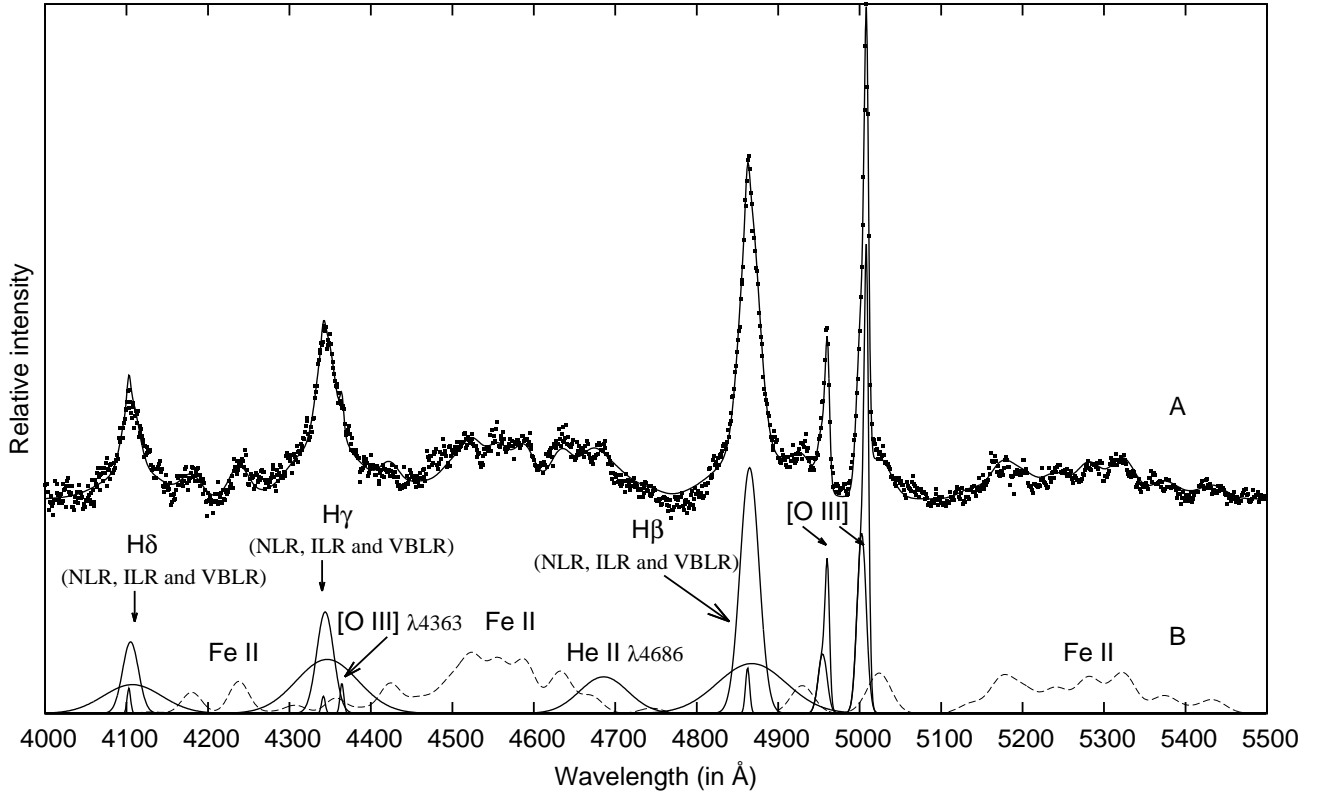}
\caption{An example of the fit of spectrum of SDSS\,J144825.09$+$355946.6 in
the $\lambda\lambda$ 4000-5500{\AA} region. A: The observed spectrum (dots) and the best
fit (solid line). B: Decomposition of the emission lines. FeII template is denoted
with dashed line. \label{fig:j2}}
 \end{figure*}

In this model, the number of the free parameters is reduced  assuming that lines or line components, which 
originate from the same emission region, have the same widths and shifts. 
Therefore, all narrow lines in considered ranges ([OIII] $\lambda\lambda$4959, 5007{\AA}, narrow 
Balmer lines, [NII] $\lambda\lambda$6548, 6583{\AA}, etc.) have the same parameters of the widths 
and shifts, since we assumed that they are originating in the Narrow Line Region (NLR). The [OIII] 
$\lambda\lambda$4959, 5007{\AA} lines are fitted with an additional component which describes the asymmetry 
in the wings of these lines, while the flux ratio of the $\lambda$4959:$\lambda$5007{\AA} is taken as 
1:2.99 \citep{dim07}. The ratio $\approx3$ is taken for [NII] doublet 
components.

The Balmer lines are fitted with three components, one narrow -- Narrow Line Region (NLR), and two broad 
-- Intermediate and Very Broad Line Region (ILR and VBLR), see  
\citet{po2004, bo2006, Bon09, hu2008}. All ILR components of the Balmer lines have the same widths and 
shifts. The same is for the VBLR components. Numerous optical FeII 
lines in the $\lambda\lambda$4000-5500{\AA} range are fitted with the FeII 
template\footnote{\url{http://servo.aob.rs/FeII\_AGN/}} presented in \citet{Kovacevic10} and \citet{sh2012}. 
In this FeII model, all FeII lines have the same widths and shifts, while relative intensities 
are calculated within the different FeII line groups, which have the same lower level of the transition 
\citep[see][]{Kovacevic10}. 
 
\begin{figure} 
 \centering
 \includegraphics[width=89mm,angle=0]{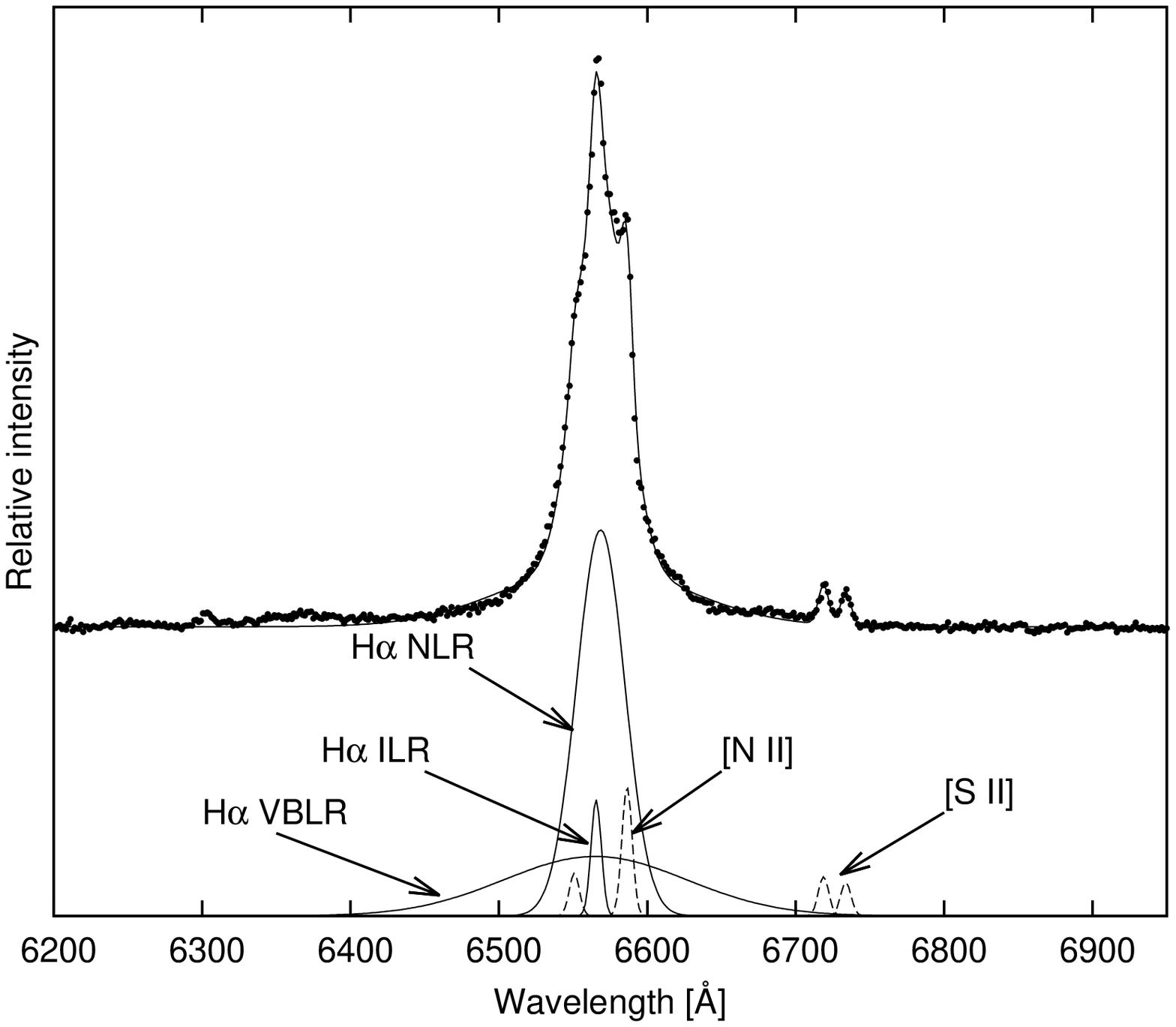}
\caption{An example of the fit of the SDSS\,J144825.09$+$355946.6 in
the $\lambda\lambda$6200-6950{\AA} region. [NII] and [SII] lines are shown with dashed 
line. \label{fig:j3}}
 \end{figure}

The detailed description of this multi-Gaussian model and the fitting procedure is given in 
\citet{Kovacevic10} and \citet{Kovacevic15}. The examples of 
 the best fit in the $\lambda\lambda$4000-5500{\AA} and $\lambda\lambda$6200-6950{\AA} ranges are 
 given in Figs.~\ref{fig:j2} and ~\ref{fig:j3}.

\subsubsection{Measuring the optical spectral parameters} \label{measuring_opt}

After performing the decomposition, we measured different 
spectral parameters of all considered optical emission lines and their components.

 Kinematic parameters, Doppler widths and velocity shifts of emission lines are directly obtained as a 
 product of fitting procedure. Additionally, we measured the FWHM of 
 broad H$\beta$ (ILR+VBLR component), FWHM(H$\beta$). The EWs of emission lines have been measured 
 with respect to pure QSO continuum (after subtraction of the host contribution) below the 
 lines \citep[see][]{Kovacevic10}. The flux of the pure QSO continuum is measured at $\lambda$5100{\AA}, and 
 continuum luminosity is calculated using the formula given in  \citet{Peebles93}, with adopted 
 cosmological parameters: $\Omega_M$=0.3, $\Omega_\Lambda$=0.7 and $\Omega_k$=0, and Hubble 
 constant $\rm H_{0}$=70 km s$^{-1}$ Mpc$^{-1}$. The mass of the black hole ($\rm M_{BH}$) is calculated 
 using the improved formula from \citet{Vestergaard06}, for the host light corrected L$_{5100}$, given 
 in \citet{Feng14}. All measured properties that we used in this work are given in Table~\ref{tab:opt}. The distributions
 of measured parameters from optical spectra are given in Fig.~\ref{fig:hist}.

\begin{table*}
{\footnotesize
\caption{The optical parameters that we measured for the sample of 82 Type 1 AGNs. A full version of 
this table is available at the electronic format. \label{tab:opt}}
\centering{
\begin{tabular}{lcccccccccc}
\hline\hline
SDSS name&EWH$\beta$N &EWH$\beta$B& EWFeII& [OIII]/H$\beta$ &FWHMH$\beta$& L5100       &EW[OIII]$^{*}$&[NII]/H$\alpha$&$\rm logM_{BH}$&F$_{H}$\\
    &{\AA}       &{\AA}      & {\AA} &                    &km s$^{-1}$ & erg s$^{-1}$&{\AA}       &                  &M$_{\odot}$  &       \\
(1) &(2)&(3)&(4)&(5)&(6)&(7)&(8)&(9)&(10) &(11)\\
\hline  
  J115341.78+461242.25& 6.677&  81.483 & 191.858&  0.447&   776.539 & 42.538& 27.877 & -0.284& 5.78& 0.27\\    
  J130359.48+534729.8 & 5.319&  82.889 & 0.0    &  1.093&   4819.368& 42.669& 105.285& -0.737& 7.44& 0.42\\    
  J153552.40+575409.50& 3.610&  76.735 & 0.0    &  1.192&   4258.708& 43.449& 86.939 & -0.581& 7.72& 0.14\\   
  J134952.84+020445.10& 7.327&  57.233 & 87.446 &  1.276&   1659.097& 42.957& 227.384& -0.333& 6.65& 0.25\\    
  J145901.36+611353.59& 6.045&  144.537& 115.841&  0.737&   6636.606& 42.491& 33.989 & -0.004& 7.62& 0.65\\
\hline
\end{tabular}}
\\
\smallskip
{\bf Notes.}
(1) SDSS name, (2) EW(H$\beta$NLR), (3) EW(H$\beta$Broad)=EW(H$\beta$ILR)+EW(H$\beta$VBLR), (4) EW(FeII), (5) log([OIII]5007/H$\beta$NLR), 
(6) FWHM(H$\beta$), (7) log(L5100)=log of luminosity of AGN at $\lambda$=5100{\AA}, multiplied 
with 5100, (8) EW[OIII]$^{*}$ is EW for both components of doublet ([OIII] $\lambda$4959{\AA} + [OIII] $\lambda$5007{\AA}), 
(9) log([NII]6583/H$\alpha$NLR), (10) log of black hole mass, log($\rm M_{BH}$), (11) Host 
fraction in the $\lambda\lambda$4160-4210{\AA} range, obtained in the decomposition. 
}
\end{table*}

\subsection{MIR properties of the AGNs} \label{MIR_properties}

\begin{figure} 
 \centering
 \includegraphics[width=89mm,angle=0]{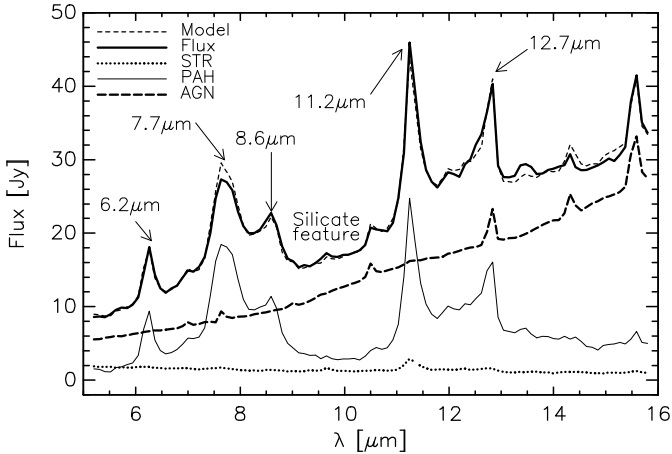}
\caption{An example of fitting of the IRS spectrum of {MRK0662}, using {\sl deblendIRS} software. 
The spectra is decomposed to the AGN, PAH and stellar component (STR). The PAH and silicate features
are marked on the graph. \label{fig:IRfit}}
 \end{figure}

To study the AGN MIR properties of the sample, we need to disentangle the AGN emission from interstellar 
PAH, and stellar (STR) components, using some of the existing tools for spectral 
decomposition of IRS data. We used {\sl deblendIRS}\footnote{\url{http://denebola.org/ahc/deblendIRS}} 
routine, written in IDL \citep{2015ApJ...803..109H}. Having a collection of real spectral templates, this 
software chooses the best linear combination of one stellar template, one PAH template, and one AGN template, 
to model an IRS spectra, only in the spectral range 5.3--15.8$\mu$m. An example of fitting of an IRS spectrum
is present on the Fig.~\ref{fig:IRfit}. For the stellar templates, the 
routine uses 19 local elliptical and S0 galaxies, the 56 PAH templates are IRS spectra of normal 
starfoming and SB galaxies at z$\le$ 0.14, while the AGN templates are 181 IRS spectra of sources
classified in the optical as quasars, Seyfert galaxies, LINERs, blazars, optically obscured
AGNs and radio galaxies.

\begin{figure} 
 \centering
 \includegraphics[width=89mm,angle=0]{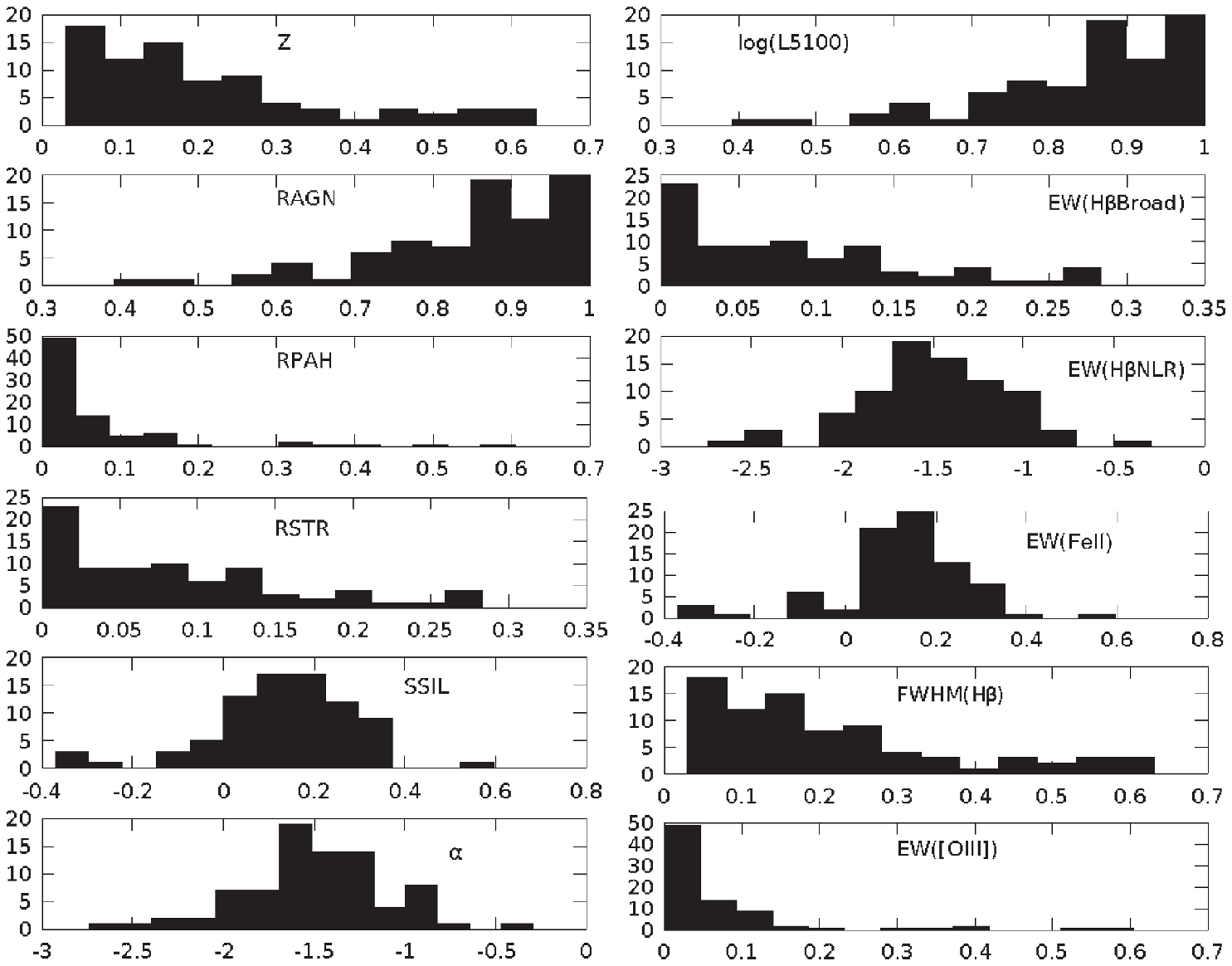}
\caption{Distribution of redshifts and main obtained parameters from fitting MIR (left) and optical 
(right) spectra (see the text).
\label{fig:hist}}
\end{figure}

The main fitting results are fractional contributions 
of AGN, PAH and stellar components to the integrated 5--15$\mu$m luminosity, named RAGN, RPAH and RSTR, 
respectively (RAGN+RPAH+RSTR=1), spectral index of the AGN component 
(assuming a power law continuum $f_{\nu} \propto \nu^{\alpha}$, between 8.1 and 12.5$\mu$m), $\alpha$ and 
the silicate strength of the best fitting AGN template, $\rm S_{SIL}$ \citep[see][]{2015ApJ...803..109H}. 
They define $\rm S_{SIL}$ as a ln(F($\lambda_{p}$)/F$_{C}$($\lambda_{p}$)), where F($\lambda_{p}$) and 
F$_{C}$($\lambda_{p}$) are the flux densities of the spectrum and the underlying continuum, at the wavelength
of the peak of the silicate feature. Other results 
of the fit are flux densities of AGN, PAH and STR components (spectra), names of the used spectral galactic templates, $\chi^2$, the 
coefficient of variation of the rms error ($\rm CV_{RMSE}$), monochromatic luminosities of the source and 
fractional contribution to the restframe of the AGN component, at 6 and 12$\mu$m. The resulting parameters 
that we use in this work are given in Table~\ref{tab:ir}, while the distributions of the main parameters is 
given on the Fig.~\ref{fig:hist}. In these histograms one can see that the AGN contribution is very dominant 
in the sample, comparing to PAH and stellar emission. Silicate feature 
is usually in the emission (positive $\rm S_{SIL}$).

From the fitting results we chose 82 successful fits, based on low reduced
$\chi^2$, $\rm CV_{RMSE}$ \citep[should be $<$0.1;][]{2015ApJ...803..109H}, and the visual inspection. 
The 16 sources were rejected due to poor fitting, from which 6 are extended sources. All these 82 objects 
are point sources except one, 0615-52345-0041. 

Independently of {\sl deblendIRS} code, we calculated EWs of PAH features at 7.7 and 11.2 $\mu$m (given in 
Table~\ref{tab:ir}), using 
STARLINK software \citep[DIPSO;][]{Howarth87} on redshift corrected IRS spectra.

\begin{table*}
{\footnotesize
\caption{The MIR parameters that we calculated and measured for the sample of 82 Type 1 AGNs. A full 
version of this table is available at the electronic format. \label{tab:ir}}
\centering{
\begin{tabular}{lccccccccccc}
\hline\hline
NED Name&RAGN &RPAH& $\rm S_{SIL}$& $\alpha$ &L6                   & L12                  &EW7.7  & EW11.2 &R$_{30/15}$&$\rm \chi^{2}_{RED}$&$\rm CV_{RMSE}$\\
        &     &    &              &          &10$^{42}$erg s$^{-1}$& 10$^{42}$erg s$^{-1}$&$\mu$m & $\mu$m &           \\
(1) &(2)&(3)&(4)&(5)&(6)&(7)&(8)&(9)&(10)&(11)&(12)\\
\hline  
  MRK\_0042    &  0.61& 0.303&  0.1  & -1.443 &6.42 & 7.34& 0.690  &  0.370   &  0.30 &0.53&0.03\\
  SBS\_1301+540&  0.71& 0.024&  0.269& -1.447 &8.99 & 7.44& $<$0.01&  0.018   &  0.97 &0.16&0.03\\
  MRK\_0290    &  0.89& 0.005&  0.061& -1.858 &3.34 & 4.24& 0.028  &  0.015   &  0.73 &1.17&0.03\\
  UM\_614      &  0.92& 0.023&  0.17 & -1.656 &1.74 & 2.15& 0.007  &  $<$0.01 &  0.89 &0.73&0.06\\
  MRK\_0836    &  0.62& 0.125&  0.304& -1.633 &6.03 & 5.90& 0.071  &  0.137   &  0.43 &0.21&0.05\\
\hline
\end{tabular}}
\\
\smallskip
{\bf Notes.}
(1) NED name, (2) and (3) -- fractional contributions of the AGN and PAH components to the MIR luminosity,
respectively (RAGN+RPAH+RSTR=1), (4) $\rm S_{SIL}$ -- strength of a silicate feature, (5) spectral index, (6) and (7) monochromatic luminosities 
of the source at 6 and 12 $\mu$m, (8) and (9) EWs of the PAH features at 7.7 and 11.2$\mu$m, 
(10) Ratio of fluxes at 30 and 15$\mu$m, (11) reduced $\rm \chi^{2}$ and (12) $\rm CV_{RMSE}$. 
}
\end{table*}

\subsection{Broad-line Balmer decrement and nuclear extinction}

As can be seen on Figs.~\ref{fig:j2} and \ref{fig:j3}, we decompose the H$\alpha$ and H$\beta$ into three components: NLR, ILR and VBLR. The narrow-line Balmer decrement, H$\alpha$$^{n}$/H$\beta$$^{n}$, is calculated as a flux ratio of H$\alpha$NLR and H$\beta$NLR, while the broad-line Balmer decrement is found as H$\alpha$$^{b}$/H$\beta$$^{b}$, where H$\alpha$$^{b}$ and H$\beta$$^{b}$ are sums of the ILR and VBLR flux components. These fluxes and ratios are given in the Table~\ref{tab:apsorption}. 

\begin{figure} 
 \centering
 \includegraphics[width=79mm,angle=0]{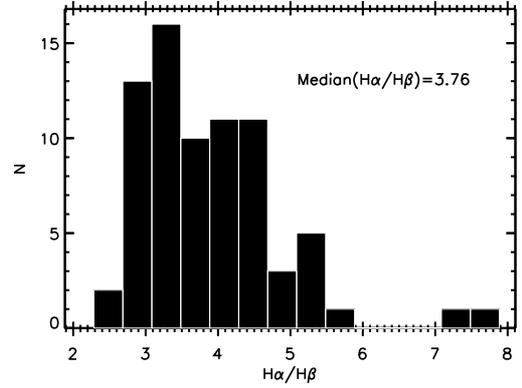}
\caption{The distribution of broad-line Balmer decrements, H$\alpha$$^{b}$/H$\beta$$^{b}$, from the Type 1 sample.}
\label{fig:Balmer}
\end{figure}

The broad-line Balmer decrement is often used for the estimation of the dust extinction in the BLR \citep{Dong08,Zhang08,Gaskell17}. The distribution of broad-line Balmer decrement is given in the histogram on the Fig.~\ref{fig:Balmer}. They are typically lower than the ones for the submm galaxies from \citet{Takata06} that have values from 5-20. Obtained H$\alpha$$^{b}$/H$\beta$$^{b}$ ratios are comparable to the values from the large sample of Seyfert 1 galaxies from \citet{Dong08}, that are well described with a log-Gaussian, with a peak at 3.05. The median value of our decrements is 3.76. However, we should note that the other effects can affect the H$\alpha$$^{b}$/H$\beta$$^{b}$ ratio, such as photoionization, recombination, collisions, self-absorption, dust obscuration, etc. \citep{Popovic03,Ilic12}.

If we use the equation (5) from \citet{Zhang08}, we could estimate the minimum color excess, E(B-V), from the sample to be -0.263. For that value of E(B-V), one can use the reddening curve from \citet{Calzetti00} to calculate the extinction, A$_{\lambda}$ at any specific wavelength.

\begin{table*}
{\footnotesize
\caption{Flux measurements of the hydrogen recombination lines and obtained Balmer decrements for the Type 1 sample. \label{tab:apsorption}}
\centering{
\begin{tabular}{cccccccc}
\hline\hline
Plate-MJD-Fiber   &H$\alpha$NLR & H$\beta$NLR &   H$\alpha$Broad & H$\beta$Broad& H$\alpha$$^{b}$/H$\beta$$^{b}$ & H$\alpha$$^{n}$/H$\beta$$^{n}$ \\
                  &10$^{-17}$erg s$^{-1}$cm$^{-2}$&10$^{-17}$erg s$^{-1}$cm$^{-2}$&10$^{-17}$erg s$^{-1}$cm$^{-2}$&10$^{-17}$erg s$^{-1}$cm$^{-2}$&   &    \\
(1) & (2)&(3)&(4)&(5)&(6)&(7)\\
\hline  
 0276-51909-0251&    1.39E+03&  3.08E+02&   34098.93&      1.09E+04&    3.13&  4.52\\
 0339-51692-0042&    7.51E+02&  1.16E+02&   7841.89 &      2.26E+03&    3.47&  6.49\\
 0340-51990-0485&    2.86E+02&  1.26E+02&   19216.24&      6.71E+03&    2.86&  2.27\\
 0355-51788-0408&    2.78E+02&  5.76E+01&   5118.76 &      1.48E+03&    3.46&  4.83\\
 0386-51788-0086&    9.49E+02&  1.86E+02&   43266.6 &      8.57E+03&    5.04&  5.10\\
\hline
\end{tabular}}
\\
\smallskip
{\bf Notes.}
(1) SDSS Plate-MJD-Fiber, (2) H$\alpha$NLR flux, (3) H$\beta$NLR flux, (4) H$\alpha$Broad flux, (5) H$\beta$Broad flux, (6) H$\alpha$$^{b}$/H$\beta$$^{b}$ -- the Balmer decrement of the broad lines, and (7) H$\alpha$$^{n}$/H$\beta$$^{n}$ -- the Balmer decrement of the narrow lines.\\
}
\end{table*}

\section{Results} \label{sec:results}
We found that in the Type 1 AGN sample there is relatively small contribution of PAH, of RPAH$<$20\%, for 
the majority of objects (see histogram in Fig.~\ref{fig:hist}). 

On the histogram on the Fig.~\ref{fig:hist}, in Type 1 AGN sample we see more silicate 
emission ($\rm S_{SIL}>$0), than absorption ($\rm S_{SIL}<$0), as mentioned in \citet{Weedman12} and 
\citet{Stalevski11}. \citet{Hao07} suggested that the QSOs are characterized by silicate emission, while 
Sy1s have equally distributed emission or weak absorption. That is in the agreement with our
findings. The silicate feature in the absorption means there is a cooler 
dust between the observer and the hotter dust responsible for the MIR continuum.

We created a linear correlation matrix for all MIR and optical parameters together. Pearson correlation
coefficients and the P-values are given in the Table~\ref{Tab5}, higher and lower, respectively. 
Significant correlations are marked with stars.

\hspace*{-8cm}  
\vspace*{22cm}
\begin{table*}
\caption{The correlation matrix for the optical and MIR parameters. The higher values are the correlation
coefficients, while the lower values are the the P values. The correlations with P$<$0.05 are marked
with *.
(2) EW(H$\beta$NLR), (3) EW(FeII), (4) log([OIII]5007/H$\beta$NLR), (5) FWHM(H$\beta$), 
(6) log of luminosity of AGN at $\lambda$=5100{\AA}, multiplied with 5100, (7) EW([OIII]) for both [OIII] components, 
(8) and (9) RAGN and RPAH -- fractional contributions of the AGN 
and PAH components to the MIR luminosity, respectively, (10) $\rm S_{SIL}$, (11) spectral index, 
(12) Monochromatic luminosity of the source at 12 $\mu$m, (13) EW of the PAH feature at 7.7$\mu$m,
(14) log of black hole mass, (15) Ratio of fluxes at 30 and 15$\mu$m, and (16) EW(H$\beta$Broad).\label{Tab5}}

\begin{footnotesize}
\rotatebox{90}{
\hspace*{-22cm}
\resizebox{20cm}{!}{%

\begin{tabular}{|r|c|c|c|c|c|c|c|c|c|c|c|c|c|c|c|} 
\hline
    & EW(H$\beta$NLR) & EW(FeII) & log([OIII]/H$\beta$) &FWHM(H$\beta$)&log(L5100)&EW([OIII])&
RAGN & RPAH & $\rm S_{SIL}$ &$\alpha$ &L12 &EW(PAH7.7) &$\rm M_{BH}$&$\rm R_{30/15}$&EW(H$\beta$Broad) \\
\hline
  (1) & (2) &(3) &(4) &(5) &(6) &(7) & (8) &(9) &(10) &(11) &(12)&(13) &(14) &(15) &(16) \\ \hline
  EW(H$\beta$NLR)&      1&          0.233*&    -0.534*&            -0.322*&  -0.447*&   0.317*&  -0.340*&  0.385*&   0.004&   -0.181&  -0.311*&   0.390*&    -0.526*&   -0.096&   -0.020   \\
  EW(H$\beta$NLR)&      --&         0.035&     2.90E-7&             0.003&    2.49E-5&  0.004&   0.002&     3.51E-4& 0.971&   0.104&    0.004&    0.002&     3.96E-7&   0.490&    0.857    \\ \hline
  EW(FeII)&             0.233*&     1&         -0.479*&            -0.324*&   -0.185&   -0.336*& -0.206&   0.315*&   -0.156&  0.119&    -0.008&   0.379*&    -0.349*&   -0.255&   0.282*   \\
  EW(FeII)&             0.035&      --&        6.10E-6&             0.003&    0.097&    0.002&   0.063&    0.004&    0.160&   0.285&    0.945&    0.003&     0.001&     0.062&    0.010    \\ \hline
  log([OIII]/H$\beta$)&-0.534*&    -0.479*&   1&                 0.300*&    0.040&    0.443*&  0.293*&   -0.262*&  -0.203&  -0.316*&  0.104&    -0.265*&   0.242*&    -0.138&   0.009    \\
  log([OIII]/H$\beta$)&2.90E-7&   6.10E-6     &--&               0.006&     0.722&    3.41E-5  &0.008&   0.018&    0.069&   0.004&    0.356&    0.039&     0.029&     0.322&    0.937    \\ \hline
  FWHM(H$\beta$)&       -0.322*&   -0.324*&    0.300*&              1&        0.153&    -0.048&  0.094&    -0.139&   0.147&   -0.029&   0.161&    0.009&     0.711*&    0.071&    0.164     \\
  FWHM(H$\beta$)&       0.003&     0.003&      0.006&               --&       0.169&    0.665&   0.400&    0.213&    0.187&   0.793&    0.149&    0.947&     7.35E-14   &0.609&   0.140    \\ \hline
  log(L5100)&         -0.447*&   -0.185&     0.040&               0.153&    1&        -0.265*& 0.549*&   -0.540*&  0.211&   0.310*&   0.663*&   -0.449*&   0.787*&    0.374*&   -0.080   \\
  log(L5100)&         2.49E-5&   0.097&      0.722&               0.170&    --&       0.016&   9.08E-8&  1.66E-7&  0.057&   0.004&    1.16E-11  &2.80E-4&  0&         0.005&    0.473    \\ \hline
  EW([OIII])&             0.317*&    -0.336*&    0.443*&           -0.048&   -0.265*&   1&       0.082&    -0.029&   -0.207&  -0.368*&  -0.177&   -0.002&    -0.182&    -0.096&   -0.037    \\
  EW([OIII])&             0.004&     0.002&      3.41E-5&           0.665&    0.016&    --&      0.464&    0.797&    0.062&   6.61E-4   &0.112&   0.985&     0.101&     0.489&    0.741    \\ \hline
  RAGN&               -0.340*&   -0.206&     0.293*&              0.094&    0.549*&   0.082&   1&        -0.823*&  0.097&   0.293*&   0.416*&   -0.688*&   0.436*&    0.292*&   -0.161   \\
  RAGN&               0.002&     0.063&      0.008&               0.400&    9.08E-8&  0.464&   --&       0&        0.385&   0.007&    1.03E-4&  8.70E-10   &4.18E-5&  0.032&    0.148    \\ \hline
  RPAH&               0.385*&    0.315*&     -0.262*&            -0.139&   -0.540*&   -0.029&  -0.823*&  1&        -0.180&  -0.304*&  -0.270*&   0.866*&   -0.458*&   -0.521*&  0.013     \\
  RPAH&               3.51E-4&   0.004&      0.018&               0.213&    1.66E-7&  0.797&   0&         --&      0.105&   0.005&    0.014&    0&         1.52E-5&   5.37E-5   &0.906   \\  \hline
  $\rm S_{SIL}$&      0.004&     -0.156&     -0.203&              0.147&    0.211&    -0.207&   0.097&   -0.180&   1&       0.228*&   0.061&    -0.393*&   0.201&     0.291*&   8.15E-4 \\
  $\rm S_{SIL}$&      0.971&     0.160&      0.069&               0.187&    0.057&     0.062&   0.385&   0.105&    --&      0.039&    0.586&    0.002&     0.070&     0.033&    0.994     \\ \hline
  $\alpha$&           -0.181&    0.119&      -0.316*&            -0.029&    0.310*&   -0.368*& 0.293*&   -0.304*&  0.228*&  1&        0.191&    -0.329*&   0.200&     0.462*&   0.010    \\
  $\alpha$&           0.104&     0.285&      0.004&               0.793&    0.004&    6.61E-4  &0.007&   0.005&    0.039&   --&       0.086&    0.009&     0.072&     4.35E-4&  0.926    \\ \hline
  L12&               -0.311*&   -0.008&     0.104&                0.161&    0.663*&   -0.177&  0.416*&   -0.270*&  0.061&   0.191&    1&         -0.162&   0.536*&    0.024&    -0.122    \\
  L12&                0.004&     0.945&      0.356&               0.149&    1.16E-11  &0.112&  1.03E-4&  0.014&    0.586&   0.086&    --&         0.213&   2.11E-7&   0.864&    0.274    \\  \hline
  EW(PAH7.7)&           0.390*&    0.379*&     -0.265*&             0.009&    -0.449*&  -0.002&  -0.688*&  0.866*&   -0.393*& -0.329*&  -0.162&   1&         -0.300*&   -0.408*&  0.115    \\
  EW(PAH7.7)&           0.002&     0.003&      0.039&               0.947&    2.80E-4&  0.985&   8.70E-10  &0&       0.002&   0.009&    0.213&    --&        0.019&     0.006&    0.378    \\  \hline
  $\rm M_{BH}$&      -0.526*&   -0.349*&    0.242*&              0.711*&    0.787*&   -0.182&  0.436*&   -0.458*&  0.201&   0.200&    0.536*&   -0.300*&   1&         0.324*&   0.065     \\
  $\rm M_{BH}$&       3.96E-7&   0.001&      0.029&               7.35E-14  &0&        0.101&  4.18E-5&  1.52E-5&  0.070&   0.072&    2.11E-7&  0.019&     --&        0.017&    0.559    \\ \hline
  R$_{30/15}$&       -0.096&   -0.255&     -0.138&               0.071&    0.374*&   -0.096&   0.292*&   -0.521*&  0.291*&  0.462*&   0.024&    -0.408*&   0.324*&    1&        0.149    \\
  R$_{30/15}$&        0.490&    0.062&      0.322&                0.609&    0.005&    0.489&   0.032&    5.37E-5&  0.033&   4.35E-4&  0.864&    0.006&     0.017&     --&       0.283    \\  \hline
  EW(H$\beta$Broad)&       -0.020&    0.282*&     0.009&                0.164&    -0.080&   -0.037&  -0.161&   0.013&    8.15E-4  &0.010&   -0.122&   0.115&     0.065&     0.149&    1         \\
  EW(H$\beta$Broad)&        0.857&     0.010&      0.937&               0.140&    0.473&    0.741&   0.148&    0.906&    0.994&   0.926&    0.274&    0.378&     0.559&     0.283&    --        \\  
  \hline
\\
\end{tabular}
}}
\end{footnotesize}
\end{table*}
\clearpage

\subsection{Expected and confirmed relations} \label{confirmed}

\begin{figure} 
 \centering
 \includegraphics[width=89mm,angle=0]{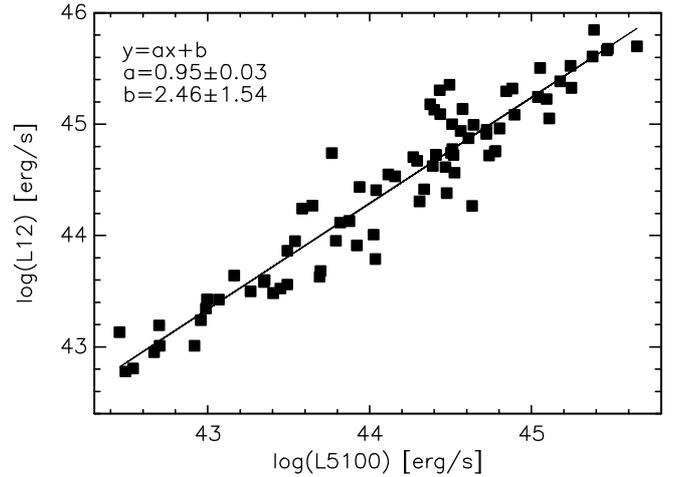}
\caption{The correlation between the AGN continuum luminosity (L5100) at the optical 
wavelengths and monochromatic luminosity of the source at 12$\mu$m. The Pearson' correlation coefficient is 
$\rho$=0.66; with P$<$ 0.00001. The linear relation between these quantities is given as a solid line
and the coefficients $a$ and $b$ are given on the plot.} \label{fig:L5100_L6}
 \end{figure}

One of the most expected results is the correlation between the AGN continuum luminosity (L5100) at 
optical wavelengths and total luminosities at 6 and 12$\mu$m, with $\rho$=0.67 and 0.66, while 
P$<$0.00001 (Fig.~\ref{fig:L5100_L6}). 
Also, there is the dependence of RAGN and RPAH with L5100, $\rho$=0.55 and -0.54, respectively, 
with P$<$0.00001, which is shown in numerous works \citep[e.g.][]{Vika17}. 

Another expected relation that we found is that RAGN 
and RPAH are in trend with redshift; Pearson' coefficients are $\rho$=0.31; 
P=0.0038 and $\rho$=-0.24; P=0.03, respectively. Finally, RAGN and RPAH are in trend with the 
luminosities at 6$\mu$m ($\rho$=0.39, P=0.0003 and $\rho$=-0.27; P=0.014) and at 12$\mu$m ($\rho$=0.41; 
P=0.0001 and $\rho$=-0.27; P=0.014). 

$\rm S_{SIL}$ is only weakly correlated with $\alpha$, with $\rho$=0.228, 
P=0.039, and that is already shown in the literature \citep{2015ApJ...803..109H,Hao07}.

\begin{figure} 
 \centering
 \includegraphics[width=89mm,angle=0]{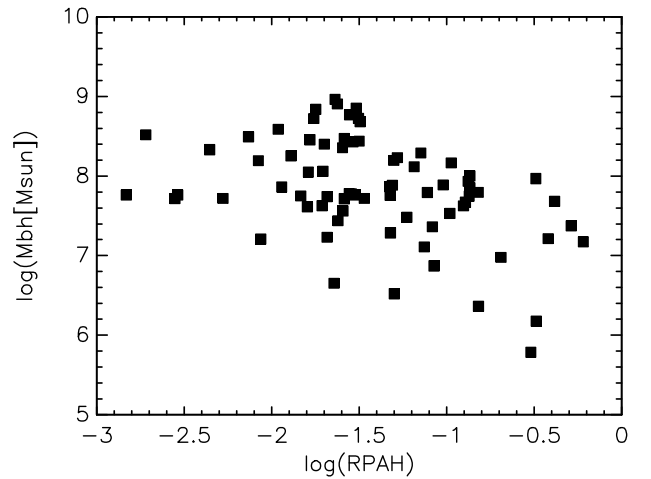}
\caption{A trend between RPAH and the black hole mass, $\rm M_{BH}$. The Pearson' correlation 
coefficient is $\rho$=-0.44; with P=0.00004.} \label{fig:Mbh_RPAH}
 \end{figure}

Among known anticorrelations that we expected to obtain is between RPAH and black hole mass, 
$\rm M_{BH}$ \citep[for example][]{Sani10}, see Fig.~\ref{fig:Mbh_RPAH}. Here we obtained a slight trend with Pearson' coefficient $\rho$=-0.44, with P=0.00004; PAH may be more dominant in AGNs with lower M$_{BH}$.

\begin{figure} 
 \centering
 \includegraphics[width=89mm,angle=0]{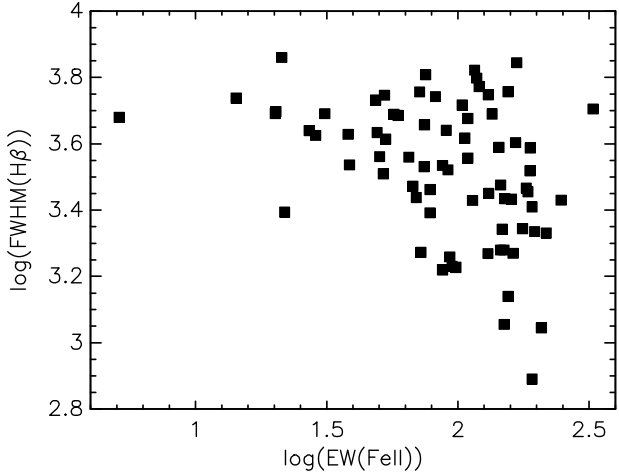}
\caption{The correlation between EW(FeII) and FWHM(H$\beta$); $\rho$=-0.32; P=0.003; a part of the EV1 
from BG92.} \label{fig:EWFeII_FWHMHb}
 \end{figure}

We obtained expected correlations between EW([OIII]) and EW(FeII) ($\rho$=-0.34; P=0.002), as well as 
between EW(FeII) and FWHM(H$\beta$) ($\rho$=-0.32; P=0.003, see Fig.~\ref{fig:EWFeII_FWHMHb}), which 
are a part of the EV1 from BG92.

\begin{figure} 
 \centering
 \includegraphics[width=89mm,angle=0]{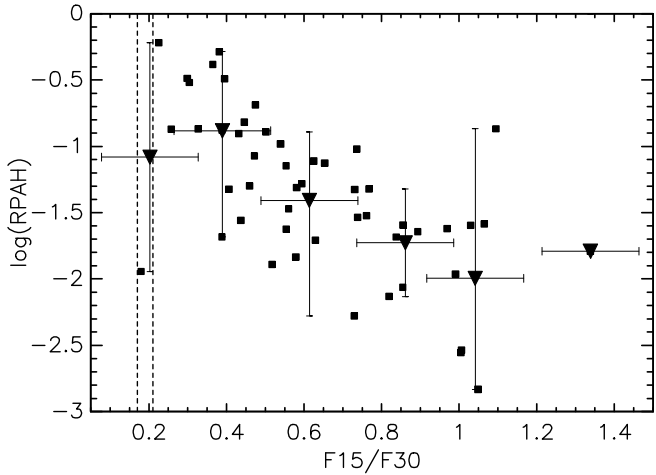}
\caption{Fractional contribution of PAH component in MIR luminosity, $\rm RPAH$, compared with ratio of fluxes
at 15 and 30 $\mu$m, $\rm R_{15/30}$. Vertical dashed lines mark the transition region between AGN and SB, according 
to \citet{Brandl06}. The Pearson' correlation coefficient is $\rho$=-0.52; with P= 0.00005. The triangles
mark the data in the bins whose widths on the x-axis are 0.25.} \label{fig:kol15_30_pah}
 \end{figure}

\subsection{Starbursts at MIR wavelengths}

At MIR wavelengths, we may estimate the SB contribution to the total radiation based on the RPAH result. 
Here, we additionally, compare the RPAH with other two usual methods
by which SB is estimated. The first is the ratio of the fluxes at 15 and 30 $\mu$m, the most accurate
method, as suggested by \citet{Brandl06}, see Fig.~\ref{fig:kol15_30_pah}.
There is a significant correlation between RPAH and this ratio. On this plot, x-axis is divided to the bins 
of the width 0.25 and binned data are shown with the triangles.

\begin{figure} 
 \centering
 \includegraphics[width=89mm,angle=0]{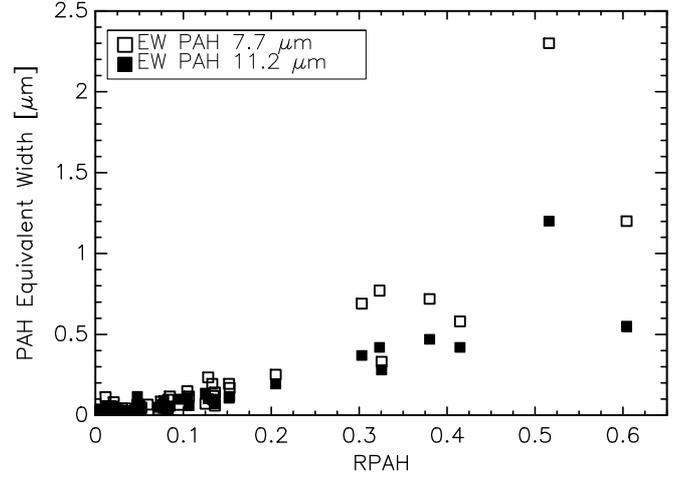}
\caption{EWs of 7.7 and 11.2$\mu$m PAH features, compared with the 
fractional contribution of PAH component, {\rm RPAH}. Pearson' correlation coefficients are $\rho$=0.87; with 
P$<$0.0001, and $\rho$=0.89; with P$<$0.0001, for PAHs at 7.7 and 11.2 $\mu$m, 
respectively. \label{fig:EW_RPAH}}
 \end{figure}

The second criterion is the the strength of
some PAH feature; here we present the EWs of the 7.7 and 11.2$\mu$m PAH features, on Fig.~\ref{fig:EW_RPAH}. There is a 
significant dependence between EWs and RPAH. As suggested 
by \citet{Lutz98a}, the objects with EW7.7 $\mu$m $>$1 are SB dominated, while the rest are AGN
dominated. On this plot, EW7.7 $\mu$m $>$1 is present only 
for two objects. Interestingly, only these two objects have RPAH $>$ 50\%.

\subsection{Comparison of the optical and MIR parameters}
In the correlation matrix (Table~\ref{Tab5}) there are several trends between the optical and MIR 
parameters. As SB significators in the MIR (RPAH, EWPAH7.7, EWPAH11.2) increase, EWs of the optical lines 
FeII and H$\beta$NLR increase, as well. On the other hand, as the MIR spectral index, $\alpha$ increases, 
a well known AGN indicator, EW([OIII]) decreases ($\rho$=-0.37, P=6.6E$^{-4}$), while the EW([OIII]) is not
related to the AGN or PAH fraction. 

We do not notice any trend of PAH EW or RPAH with the FWHM(H$\beta$) line, as suggested by \citet{Sani10}, 
who found that narrower broad H$\beta$ lines have a stronger PAH emission in the Type 1 AGNs. However, our 
later analysis (Section~\ref{PCA},~\ref{PCAadd}), will show that there is a connection 
between RPAH and FWHM(H$\beta$).

Considering this comparison it should be emphasized that the BPT and MIR SB/AGN diagnostics do not necessarily trace the contribution of an AGN to the total power of the galaxy. Therefore, there may exist some other effects which can affect one or both diagnostics.

\subsection{Comparison between starburst fraction at optical (BPT diagram) and at MIR wavelengths (RPAH)} \label{sec:CompareBPT}
\begin{figure} 
 \centering
 \includegraphics[width=89mm,angle=0]{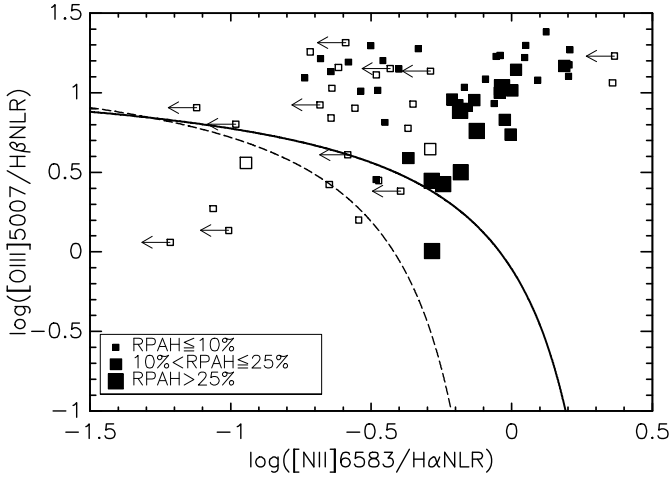}
\caption{BPT diagram for Type 1 AGN sample where RPAH contribution is illustrated with the symbol size (see the legend). We show
only the objects where both narrow H${\alpha}$ and [NII] lines are not zero. Full squares
represent the objects where narrow lines H$\alpha$ and [NII] are well detected. Empty squares mark objects
where [NII] doublet can not be well resolved within broad H$\alpha$ line, therefore these fits are less 
confident (see the Appendix~\ref{a1}). The arrows mark upper limits of log([NII]6583/H$\alpha$NLR). The 
dashed line is the separation curve from \citet{Kauffmann03}, while the solid line is the separation curve 
from \citet{Kewley01}. 
\label{fig:BPT}}
 \end{figure}

Traditionally, BPT diagram have been used for optical diagnostics between AGN, composites and SB 
\citep{Baldwin81,Kewley01}. In Fig.~\ref{fig:BPT} we show the BPT diagram of the Type 1 data sub-sample of 
69 objects with available range of 6200-6550{\AA} which covers H$\alpha$ and [NII] lines. To find the real 
ratio between the AGN and PAH contribution in the MIR spectra and present it on BPT diagram, we excluded 
stellar contribution, by using the formula RPAH=RPAH/(RPAH+RAGN) from now on. On the diagram, the RPAH is 
quantified by the three different symbol sizes. Clearly, there are a couple of objects with a low RPAH,
that lie below the solid separation curve from \citet{Kewley01}; they should be above that
curve, by the optical diagnostics. Similarly, a few SB dominated objects, 
according to the MIR fitting, lie on the AGN part of BPT diagram. These results suggest that there might 
be a significant difference between optical and MIR SB quantification.

\begin{figure} 
 \centering
 \includegraphics[width=89mm,angle=0]{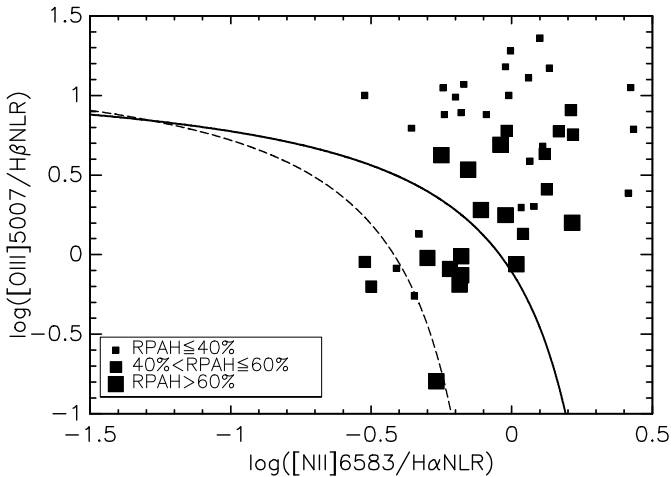}
\caption{BPT diagram for Type 2 AGN sample where RPAH contribution is illustrated with the symbol size (see 
the legend). The data are taken from the literature (Table~\ref{tab:lit}). The separation curves are the same
as in Fig.~\ref{fig:BPT}.
\label{fig:BPT_RPAH_lit}}
 \end{figure}

To confirm these doubts, taking into account that it is complicated to decompose the narrow lines in the 
Type 1 AGNs \citep[see the Appendix~\ref{a1} and][]{Popovic11}, we chose a sample of Type 2 AGNs (see 
Table~\ref{tab:lit} and Section~\ref{mixed}). We made another BPT diagram, composed from the various LINER, 
HII regions and Seyfert 2 galaxies, from the samples of \citet{2015ApJ...803..109H} and \citet{Garcia16}. 
That BPT diagram is shown in the Fig.~\ref{fig:BPT_RPAH_lit}. Again, the symbol size represents RPAH value. 
Similarly as above, we notice that often these optical and MIR results give different information about SB and AGN ratio.

\begin{figure} 
 \centering
 \includegraphics[width=89mm,angle=0]{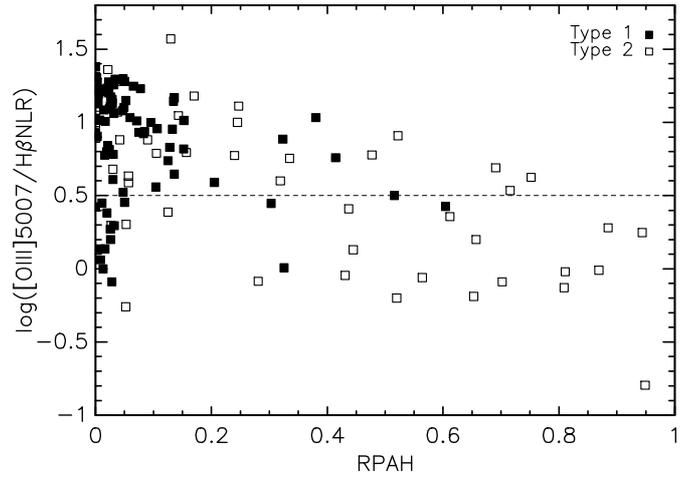}
\caption{Comparison between RPAH (PAH fraction from MIR data) with log(${\rm [OIII]5007/ H \beta}$NLR) 
(from optical data), for the sample of Type 1 AGNs (Pearson' correlation coefficient is $\rho$=-0.26; P=0.018) -- 
full squares, and for the sample of Type 2 AGNs ($\rho$=-0.60; P=4.77$\times$ 10$^{-6}$) -- 
open squares. \label{fig:R_RPAH}}
 \end{figure}

\citet{Popovic11} suggested that one axis of the BPT diagram, the ratio 
R=log(${\rm [OIII]5007/ H \beta}$NLR), may be the significant 
indicator of the SB activity, where the objects with R$<$0.5 are SB dominated and the rest are AGN
dominated. Furthermore, these authors showed that these two groups 
(R$<$0.5 and R$>$0.5) belong to the different populations of objects, since they show certain different 
optical characteristics, and that is the essence of the BPT diagram. Therefore, we 
compare the RPAH (from MIR data) and this ratio $\rm R$, on the Fig.~\ref{fig:R_RPAH} for
both Type 1 and Type 2 samples. Here we found quite weak trend of $\rho$=-0.26 and P=0.018
for Type 1 and stronger $\rho$=-0.6 and P=4.77$\times$ 10$^{-6}$ correlation for the Type 2 sample. This 
graph shows that the objects with R$<$0.5 are not always SB dominated (based to MIR data), hence there 
may be some other reason why they are different at optical wavelengths, or the origin of the optical and 
MIR radiation may be different. It can be seen in Fig.~\ref{fig:R_RPAH} that, in general, the ratio
log(${\rm [OIII]5007/ H \beta}$NLR) is decreasing as RPAH is increasing.

One should be aware that this comparison between MIR and optical diagnostics is limited by the factors such as the used data, radiation mechanisms, observed wavelength ranges and power distribution, and therefore these conclusions should not be taken literally.

\subsection{Principal Component analysis of the spectral parameters} \label{PCA}
The dependence between all calculated optical and MIR parameters is complex and therefore we used the PC 
analysis to understand the most important connections. Having our correlation matrix (Table~\ref{Tab5}), 
we chose several parameters which should contain a potentially unique information. The optical parameters 
are: EW([OIII]), EW(FeII), EW(H$\beta$NLR), EW($\rm H\beta_{broad}$) (ILR+VBLR), 
log([OIII]/H$\beta$NLR), FWHM(H$\beta$) and log($\rm L5100$), which are chosen to be 
compared with correlations found in the EV1 of BG92. The MIR 
parameters taken for analysis are: RPAH, $\alpha$ and $\rm S_{SIL}$ (since it is a possible indicator of the 
AGN geometry or inclination). 

The aim of this analysis is:
\begin{enumerate}
 \item to check possible connections between BG92 EV1 correlations (between EW(FeII) vs. EW([OIII]) 
 and EW(FeII) vs. FWHM(H$\beta$)) and some MIR spectral properties, which could give us some insight in 
 physical cause of these correlations. 
 \item to compare the RPAH with log([OIII]5007/$\rm H\beta$NLR), in order to clarify the previous results (see 
 Section~\ref{sec:CompareBPT}) about their inconsistency.
\end{enumerate}

\begin{table}
\begin{center}
\caption{PCA of the optical and MIR properties in sample of 82 AGN Type 1. \label{t02}}
\begin{tabular}{|c|c c c c|}
\hline
\footnotesize{}&Comp.1&Comp.2&Comp.3 &Comp.4   \\
\hline
\vspace*{-0.1cm} Standard deviation    & 1.617 &1.480 &1.128&1.037 \\
\vspace*{-0.1cm}Proportion of Variance &0.262 &0.219& 0.127&0.107\\
\vspace*{-0.1cm}Cumulative Proportion  &0.262& 0.481& 0.608&0.715\\
 \hline
 \vspace*{-0.1cm}RPAH &        0.447& -0.095& -0.160& -0.109\\
 \vspace*{-0.1cm}$\alpha$&    -0.140&  0.479&  0.101&  0.087\\
\vspace*{-0.1cm} $\rm S_{SIL}$ &       -0.149&  0.274&  0.262& -0.677\\
\vspace*{-0.1cm}EW([OIII]) &     0.041& -0.530&  0.228& -0.079 \\
\vspace*{-0.1cm}EW(FeII)    &     0.370&  0.314& -0.348&  0.231 \\
\vspace*{-0.1cm}EW(H$\beta$NLR)    &    0.472& -0.052&  0.321& -0.304 \\
\vspace*{-0.1cm}EW($\rm H\beta_{broad}$) &    0.057&  0.059& -0.671& -0.280 \\
\vspace*{-0.1cm}log([OIII]/H$\beta$NLR) &-0.349 &-0.466& -0.192&  0.149      \\
\vspace*{-0.1cm}FWHM(H$\beta$)   &     -0.315& -0.092& -0.346& -0.494   \\
\vspace*{-0.1cm}log(L5100)    &     -0.415&  0.274&  0.108&  0.160   \\
\hline
\end{tabular}
\end{center}
\end{table}

For the PCA we used the task {\sl princomp} with the option {\sl cor=true}, in {\sl R}. The results 
of the PCA are shown in the Table~\ref{t02}. The first four components together account for 71\% of 
the variance. The first two components are the most important underlying parameters that govern 
the observed properties of the AGN sample. Both, the first and the second eigenvector account for 
$\approx20\%$ variance, while each of the next two eigenvectors account for $\approx10\%$ variance.

\begin{figure} 
 \centering
 \includegraphics[width=89mm,angle=0]{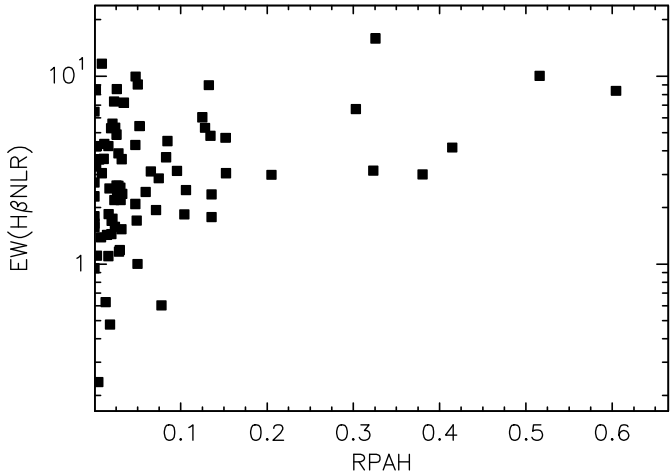}
\caption{Comparison between RPAH (PAH fraction from MIR data) with EW(H$\beta$NLR). Pearson' 
correlation coefficient is $\rho$=0.38; P=0.00035.  \label{fig:EWHbNLR_RPAH}}
 \end{figure}

The PCA indicates that the first principal component is dominated by RPAH and EW(H$\beta$NLR).
While RPAH, EW(H$\beta$NLR) and EW(FeII) have positive projections on the first eigenvector, 
log([OIII]/H$\beta$NLR), FWHM(H$\beta$) and log$\rm (L5100)$ have negative projections. 
Therefore, this eigenvector indicates that as the SB contribution, measured in IR, is stronger, 
EW(H$\beta$NLR), EW(FeII) increase, but the optical luminosity, FWHM(H$\beta$) and log([OIII]/H$\beta$NLR) 
ratio decrease. It seems that the strongest indicator of SB presence in optical is EW(H$\beta$NLR)
(see Fig.~\ref{fig:EWHbNLR_RPAH}), and consequently, the ratio of log([OIII]/H$\beta$NLR) is 
related to SBs as well. Note that the EW([OIII]) lines are not dependent on a SB presence (RPAH). This 
eigenvector shows that stronger SBs are present in AGNs with lower FWHM(H$\beta$) and stronger EW(FeII). 
This is one of the anticorrelations 
from BG92 EV1 (EW(FeII) vs. FWHM(H$\beta$); see Fig.~\ref{fig:EWFeII_FWHMHb}). This confirms the result 
of \citet{Sani10}, who found 
that SB presence is stronger in AGNs with lower width of the broad H$\beta$ lines.

The second eigenvector is dominated with MIR spectral index, $\alpha$ and EW([OIII]). $\alpha$ and 
EW(FeII) have positive projections, as well as $\rm S_{SIL}$ and log$\rm (L5100)$, but weaker. On the 
other hand, EW([OIII]), and consequently log([OIII]/H$\beta$NLR), have the negative projections. EW(H$\beta$NLR) 
is not projected on this eigenvector. This implies that as the $\alpha$ grows, the 
EW([OIII]) and log([OIII]/H$\beta$NLR) decreases, but EW(FeII) increases. In this eigenvector 
is projected the anticorrelation between EW([OIII]) and EW(FeII) from BG92 EV1, which seems to be 
related with $\alpha$.

The third eigenvector is dominated by EW($\rm H\beta_{broad}$), which correlates 
with the FWHM(H$\beta$) broad line and anticorrelates with EWs of the narrow lines. 

The fourth eigenvector is strongly dominated with the strength of the silicate feature ($\rm S_{SIL}$) 
which is correlated with FWHM(H$\beta$). This 
implies the connection between the widths of the broad lines and inclination angle or geometry of torus.

\section{Discussion} \label{sec:discuss}

\subsection{BPT diagram at MIR wavelengths} \label{sec:discussbpt}
In the Section~\ref{sec:CompareBPT}, we found a certain disagreement between optical and 
MIR quantifying of the SB contribution to the AGN spectra, for both the Type 1 and Type 2 AGNs. The 
presence of the Type 1 AGNs on the SB part of the BPT diagram that we noticed here is earlier 
observed \citep{Popovic09,Popovic11,Wang06}. 

We do not completely understand the reason of the 
misplacement of these objects on the BPT diagram. The extinction at the optical wavelengths 
is one of the most often explanations, although some authors believe that the radiation may come from the 
different regions and/or that the slit difference between SDSS and IRS might contribute to this 
disagreement \citep{Vika17}. We can not exclude the possibility of the imperfection of some of the methods,
such as the decomposition or fitting. These results remind to the results of 
\citet{Goulding09} and \citet{Dixon11}, who found that there may exist AGNs in half of the luminous 
IR galaxies without any evidence of AGN at the near-infrared and optical wavelengths. 

It seems that the Type 1 AGNs have a lower disagreement than the Type 2. The possible reasons for that are:
1) Maybe since the AGN signature is more prominent in the Type 1 sample; 2) The column density should be lower 
for the Type 1 than for Type 2 AGNs, thus the probability to observe the different regions of the AGN 
in optical and MIR is lower; 3) Because the Type 1 sample is more homogeneous and may be more accurate. 

Another cause of this disagreement between optical and MIR SB/AGN dominance could be the difference in the emission from within the optical or MIR wavebands sampled by the data. The only way to properly estimate which power source dominates the emission in galaxy is to sample the full SED or to apply bolometric corrections to the data. Methods of this type have been done in analyses of ULIRGs from various groups \citep{Veilleux09,Armus07,Petric07}.

\subsection{Comparison between the optical and MIR SB/AGN diagnostics} \label{sec:discussother}

As we mentioned, the objects with R$<$0.5 have somewhat different certain optical 
characteristics \citep{Popovic11,Kovacevic15}; which is believed to be caused by SB presence. However, as 
we obtained, on Fig.~\ref{fig:R_RPAH}, objects with R$<$0.5
do not always have a high RPAH contribution, therefore there may exist some other reason why these objects 
have special optical characteristics. 

PCA of the optical and MIR parameters confirms the results of BG92 and the other authors. It 
shows that R is probably 
influenced by more different physical properties of AGNs. Namely, EW($\rm H\beta$NLR) is correlated with 
SB strength in MIR, while [OIII] lines are 
correlated with $\alpha$. This means that R is indeed influenced by the
SB presence, but also influenced by some other physical property which affects $\alpha$, which may be 
the cause of disagreement between MIR and optical diagnostics of the SB presence.

\subsection{Connection between BG92 EV1 and MIR properties} \label{PCAadd}

Since BG92 established the set of correlations between AGN Type 1 spectral properties (EV1 in their PCA), 
it has been many attempts to explain their physical origin. The most frequently proposed governing mechanisms 
are: 1) Eddington ratio, L/$\rm L_{edd}$ \citep{Boroson02,Grupe04} 2) AGN orientation \citep{Bisogni17}, 3) 
and combination of these two properties \citep{Marzaini01,Shen14}.

It was proposed that BG92 EV1 correlations can be considered as a surrogate "H--R Diagram" for Type 1 AGNs, 
with a main sequence driven by Eddington ratio convolved with line-of-sight orientation 
\citep{Sulentic00,Shen14}, for a review see \citet{Sulentic15}. Also, the BG92 EV1 is considered as an 
indicator of the AGN evolution \citep{Marziani03,Grupe04,Wang06,Popovic11}. The evolution of AGNs is probably 
related with SB regions, assuming that there is stronger presence of the SB nearby the 
central engine of AGN in an earlier phase of AGN evolution, while in the later phases, the SB 
contribution probably becomes weaker and/or negligible \citep{Hopkins05,Lipari06,Wang08,Sani10}.

To understand the physical background of BG92 EV1 correlations, in Section~\ref{PCA}, we performed PCA 
using several optical and MIR spectral parameters. When interpreting the PCA results, it is important to 
take into account that an eigenvector is always specific to a certain sample, depending which observed 
parameters have been used and on the range of the parameters \citep{Grupe04}. Therefore, each individual 
sample has its own eigenvectors, e.g. when using a set of different spectral parameters for PCA, BG92 EV1
can be projected on some other eigenvector, or divided into two or more eigenvectors. In this
analysis, RPAH is chosen as 
a SB indicator \citep[see][]{Peeters04,Brandl06,Houck07}, $\alpha$ as a MIR spectral 
index of the pure AGN continuum, and $\rm S_{SIL}$ should be an indicator of the 
geometry and inclination \citep{Hao07}. The results of the PCA are summarized in Table~\ref{novaPCA}, where 
correlations between the optical and MIR parameters are denoted with uprising arrows, anti-correlations with 
decreasing arrows, and lack of any connection (projections to eigenvectors$<$0.25) with zero. In cases of 
weak connection (0.25$<$projections to eigenvector$<$0.30) we note "weak" in Table~\ref{novaPCA}. These 
results imply that the two the most interesting BG92 EV1 anti-correlations (EW(FeII) vs. FWHM(H$\beta$) and 
EW(FeII) vs. EW([OIII])) are related 
with different MIR parameters, since they are projected into two different
eigenvectors in our analysis. The first dominated with RPAH and the second with $\alpha$. The EW(FeII) 
vs. FWHM(H$\beta$) anti-correlation is connected with SB presence (RPAH), while EW(FeII) vs. EW([OIII]) seems to 
be connected with some physical property, reflected in $\alpha$. The $\alpha$ is a complex parameter which 
reflects MIR SED, and therefore depends on several physical properties as: 
i) accretion disc radiation (which depends on Eddington ratio, L/$\rm L_{edd}$, \citep[see][]{Zhang08}, 
ii) inclination, and iii) torus physical properties as geometry, dust distribution, optical depth, etc. 
\citep{Stalevski12}. 

\begin{table}
\begin{center}
\caption{The summary of the PCA performed in the Section~\ref{PCA} and discussed in 
Sections~\ref{sec:discussother} and~\ref{PCAadd}. \label{novaPCA}}
\begin{tabular}{|c|c|c|c|}
\hline
     & RPAH & $\alpha$ & $\rm S_{SIL}$\\
     & EV1  & EV2      & EV3          \\
\hline
EW([OIII])                  &0             &$\searrow$     &0               \\ \hline
EW(FeII)                    &$\nearrow$    &$\nearrow$     &0              \\ \hline
EW(H$\beta$NLR)             &$\nearrow$    &0              &$\nearrow$      \\  \hline  
EW(H$\beta$Broad)           &0             &0              &$\nearrow$ weak     \\   \hline
log(OIII5007/H$\beta$NLR)   &$\searrow$    &$\searrow$     &0                \\  \hline
FWHM(H$\beta$)              &$\searrow$    &0              &$\nearrow$      \\  \hline
log(L5100)           &$\searrow$    &$\nearrow$ weak&0               \\  \hline
\end{tabular}
\end{center}
\end{table}

Although it is supposed that they are dominantly driven by different physical properties, the relation 
between RPAH, $\alpha$ and $\rm S_{SIL}$ seems to be complex. \citet{2015ApJ...803..109H} tested the validity of the 
MIR spectral decomposition of {\sl deblendIRS} code (see their Section~3.2), and found that 
$\alpha$ is in significant correlation with nuclear spectral index derived from ground-based observations. 
Although 
$\alpha$ should be a pure AGN property, it is in a weak anticorrelation with the RPAH, EWPAH7.7 and 
EWPAH11.2 in our sample (see Table~\ref{Tab5}). The trend between $\alpha$ and RPAH may be caused 
by the reverse dependence between RPAH and RAGN (see Fig.~\ref{fig:hist}). The trend between the $\alpha$ 
and EW of PAHs is probably present because EWs are 
measured relative to the total continuum flux, with AGN flux included. We found a weak trend between $\alpha$ 
and $\rm S_{SIL}$ in our correlation matrix (Table~\ref{Tab5}) and Section~\ref{confirmed} 
\citep[as noticed by][]{Hao07,2015ApJ...803..109H}.

\citet{Wang06} used the MIR color $\alpha$(60,25) as an indicator 
of the SB presence, and found its correlation with BG92 EV1 correlations. Note, that the MIR color 
$\alpha$(60,25) is measured using the total (AGN+SB) flux, and therefore it contains information about the 
SB presence and AGN continuum slope, which are in this work separated in two parameters, RPAH and $\alpha$. 
Our results are consistent with results of \citet{Wang06}, but we give more detailed 
insight in origin of BG92 EV1 correlations.

\section{Conclusions} \label{sec:conclussion}

Here we investigate the optical and MIR spectral properties of a sample of 82 Type 1 AGNs. Additionally,
to check the results based on the narrow lines (which are superposed with broad lines in the Type 1 AGNs),
we considered a sample of 49 Type 2 AGNs. We carefully fit the optical spectra using methods described in
\citet{po2004}, \citet{Kovacevic10} and \citet{Kovacevic15}. For fitting of MIR data we used 
{\sl deblendIRS} code described in \citet{2015ApJ...803..109H}. Concerning our investigation, we can 
outline following conclusions:

\begin{enumerate} 
\item In sample of Type 1, we see more silicate emission, than absorption, which is
expected, according to \citet{Hao07,Stalevski11} and \citet{Weedman12}.
\item We did not find any linear trend in the correlation matrix between EW(PAH) or RPAH with FWHM(H$\beta$)
(see Table~\ref{Tab5}), but PCA shows anticorrelation between these properties, which is in an agreement with the result of 
\citet{Sani10}, who found that the narrower broad FWHM(H$\beta$) have stronger a PAH emission.
\item The separation between AGN and SB based on the BPT diagram does not give the same result as the 
one from MIR spectra in both Type 1 and Type 2 AGN samples. Some of the possible reasons 
are extinction at optical wavelengths \citep{Dixon11}, different sizes of slits, or the radiation may come from the 
different regions \citep{Vika17}.
\item The weak correlation between the main optical (log([OIII]5007/H$\beta$NLR)) and MIR (RPAH) 
starburst estimators
implies that the difference in the optical characteristics in the objects with log([OIII]5007/H$\beta$NLR)$<$0.5 
and $>$0.5 \citep{Popovic11,Kovacevic15} may have a different reason than the star formation 
presence.
\item PCA shows that the anticorrelations between EW(FeII) vs. FWHM(H$\beta$), as well 
as EW(FeII) vs. EW([OIII]), from BG92 EV1, probably have a different governing mechanism: the former 
is connected with SB presence, while the latter is more connected with the MIR AGN spectral index.
\item PCA implies that the ratio log([OIII]5007/H$\beta$NLR) is indeed influenced by the
starburst presence, but also influenced by some other physical property (MIR AGN spectral index), which may be 
the cause of disagreement between MIR and optical diagnostics of the starburst presence in AGN spectra.
\item A well known AGN indicator, EW([OIII]) is related to the MIR spectral index $\alpha$, but not related 
to the AGN or PAH fraction.
\item Since the BPT and MIR SB/AGN diagnostics do not necessarily trace the contribution of an AGN to the total power of the galaxy, the dissagrement between the two methods is not overly unexpected.
\end{enumerate}
Finally, here we confirm some correlations between the optical and IR spectral properties that
have been governed by presence of the SB contribution. EW(FeII) and EW(H$\beta$NLR) are correlated with RPAH.
Anticorrelation EW(FeII) vs. FWHM(H$\beta$) may be also connected with the RPAH, since they are 
projected on the same eigenvector as RPAH.

\section*{Acknowledgments}

This work is part of the project (146001) "Astrophysical Spectroscopy
of Extragalactic Objects" supported by the Ministry of Science of
Serbia.

Funding for SDSS-III has been provided by the Alfred P. Sloan Foundation, the Participating Institutions, 
the National Science Foundation, and the U.S. Department of Energy Office of Science. The SDSS-III web site 
is http://www.sdss3.org/.

SDSS-III is managed by the Astrophysical Research Consortium for the Participating Institutions of the 
SDSS-III Collaboration including the University of Arizona, the Brazilian Participation Group, Brookhaven 
National Laboratory, Carnegie Mellon University, University of Florida, the French Participation Group, 
the German Participation Group, Harvard University, the Instituto de Astrofisica de Canarias, the Michigan 
State/Notre Dame/JINA Participation Group, Johns Hopkins University, Lawrence Berkeley National Laboratory, 
Max Planck Institute for Astrophysics, Max Planck Institute for Extraterrestrial Physics, New Mexico State 
University, New York University, Ohio State University, Pennsylvania State University, University of 
Portsmouth, Princeton University, the Spanish Participation Group, University of Tokyo, University of 
Utah, Vanderbilt University, University of Virginia, University of Washington, and Yale University.

The Cornell Atlas of Spitzer/IRS Sources (CASSIS) is a product of the Infrared Science Center at Cornell 
University, supported by NASA and JPL.

Much of the analysis presented in this work  was done with TOPCAT 
(\url{http://www.star.bris.ac.uk/\~mbt/topcat/}), developed by M. Taylor.

We thank dr Ching Wa Yip, dr Nata\v{s}a Bon, dr Marko Stalevski, dr Predrag Jovanovi\'{c} and dr Giovanni Lamura for help with important issues in this work. We also thank the referee for helpful and constructive suggestions.

\appendix

\section{Estimation of the narrow emission lines for the BPT diagram} \label{a1}

The axes in the BPT diagram are the ratios of the particular narrow lines ([OIII]/H$\beta$NLR and 
[NII]/H$\alpha$NLR) and therefore the accurate measurements of the fluxes of these lines is very important 
for correct AGN/SB diagnostics.

In the Type 1 AGNs, these narrow lines overlap with the broad lines, and in some cases it is very difficult 
to distinguish them. We use the same fitting procedure as in \citet{Popovic11}, where the confidence of 
the narrow H$\beta$ component estimation and non-uniqueness in the solutions is tested and discussed 
\citep[see Appendix A in][]{Popovic11}.

As it is explained in Section~\ref{model_line}, we use the same parameters for widths and shifts for all 
considered narrow lines ([OIII], H$\beta$NLR, [NII] and H$\alpha$NLR). In this way, we are getting less 
degree of freedom in the fitting procedure and more confident fits in the case when one of these lines can not 
be distinguished well from the broad lines.

\begin{figure} 
 \centering
 \includegraphics[width=95mm]{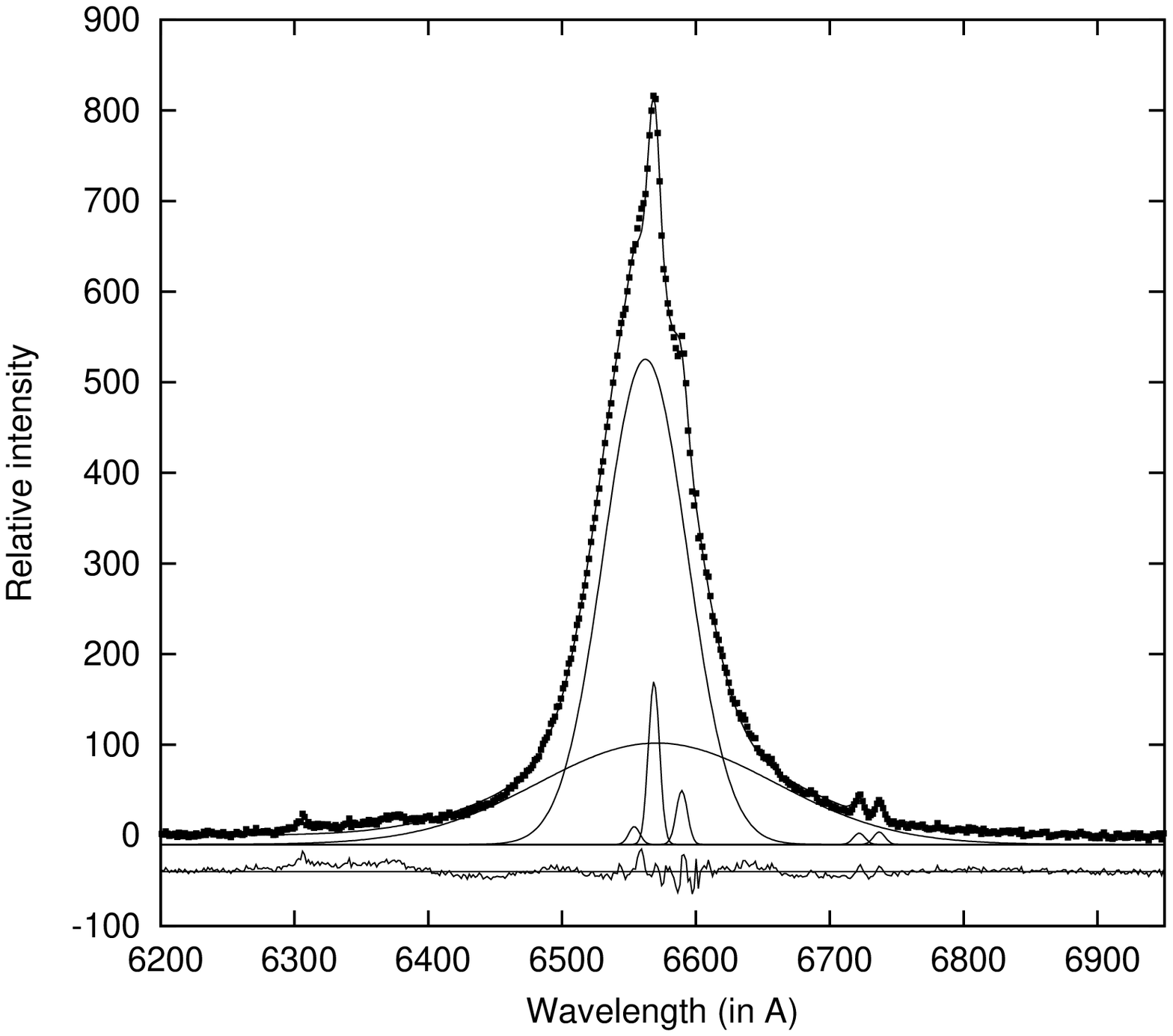}
\caption{An example of the fit of object (SDSS J130947.00$+$081948.24) with weak [NII] lines, which are 
difficult to be resolved in the broad H$\alpha$ profile. \label{fig:a1}}
 \end{figure}

\begin{figure} 
 \centering
 \includegraphics[width=89mm]{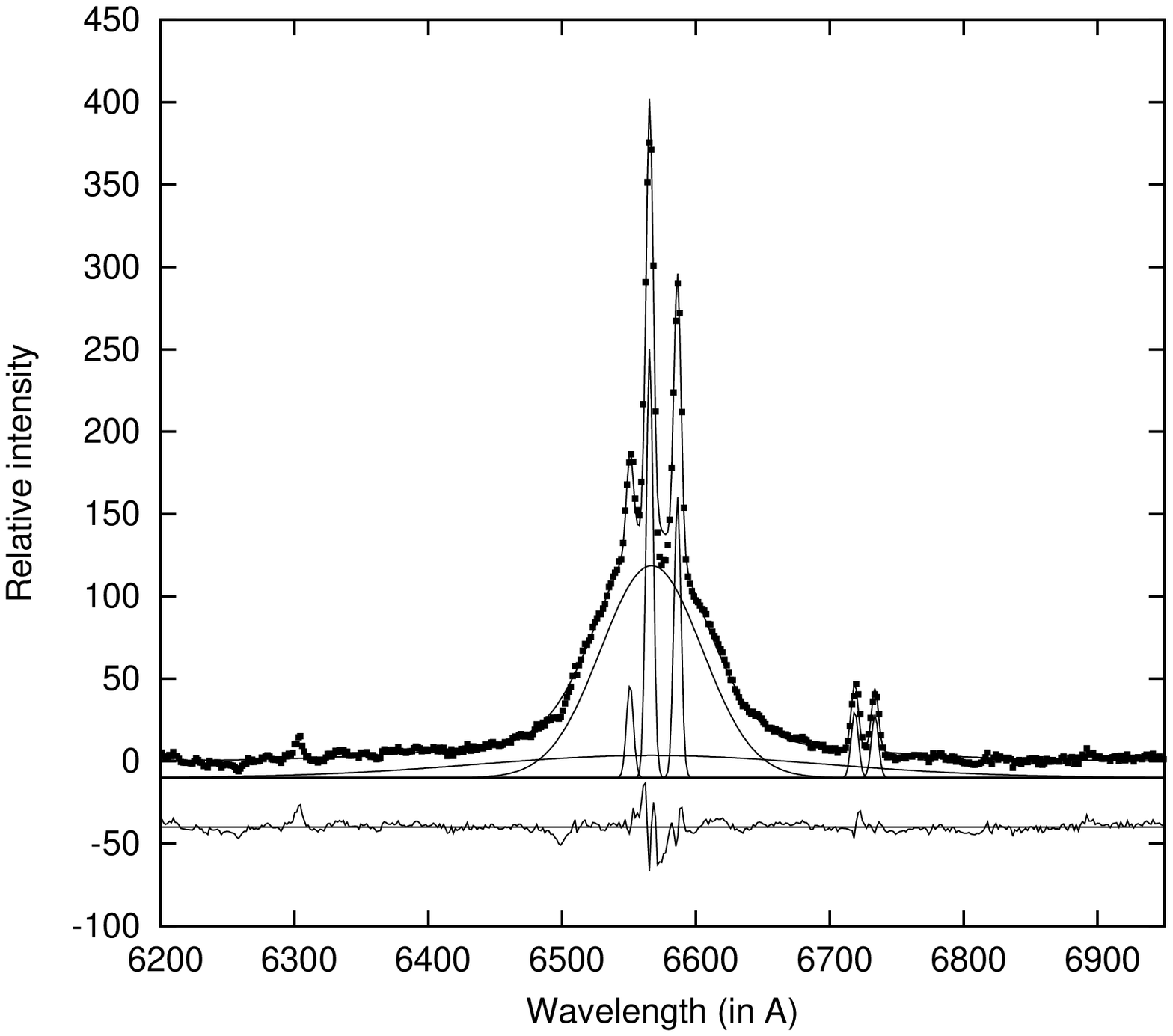}
\caption{An example of the fit of object (SDSS J135406.43$+$232549.40) with the strong and prominent [NII] 
lines. \label{fig:a2}}
 \end{figure}

In our sub-sample of 69 objects with available optical range of $\lambda\lambda$6200-6950{\AA} (which covers 
[NII] and H$\alpha$), in 26 objects [NII] lines are are very weak and barely seen in the broad H$\alpha$ 
profiles. These objects are assigned in the BPT diagram as less confident (empty squares, see 
Fig.~\ref{fig:BPT}). The example of this kind of objects is shown in Fig.~\ref{fig:a1}. On the other hand, 
in 43 objects from the sub-sample, [NII] lines can be well resolved from the broad H$\alpha$ profile (see the 
example in Fig.~\ref{fig:a2}), and these object are assigned in the BPT 
diagram as more confident (full squares in Fig.~\ref{fig:BPT}).

\bsp
\label{lastpage}
\end{document}